\documentclass[draftcls,onecolumn,10pt]{IEEEtran}
\usepackage{etex}
\usepackage{hyperref}
\usepackage{times}
\usepackage{latexsym,amsfonts,amsbsy,amssymb}
\usepackage{amsmath}
\usepackage{graphicx}
\usepackage{cite}
\usepackage{algorithm}
\usepackage{algorithmic}
\usepackage{comment}
\usepackage{multirow}
\usepackage{amsmath}
\usepackage{enumerate}
\usepackage{tabularx,booktabs}
\usepackage{xcolor}
\usepackage[curve]{xypic}
\usepackage{subfig}
\usepackage{caption}
\usepackage{color}
\usepackage{tikz}
\usetikzlibrary{automata,positioning,shapes,arrows}
\newtheorem{theorem}{Theorem}

\newtheorem{proposition}{Proposition}
\newtheorem{definition}{Definition}

\newtheorem{remark}{Remark}

\newtheorem{example}{Example}
\newtheorem{problem}{Problem}

\usepackage{float}

\begin{document}
\title{Coordination and Control of Distributed Discrete Event Systems under Actuator and Sensor Faults
\thanks{The partial support of the National Science Foundation (Grant No. CNS-1446288, ECCS-1253488, IIS-1724070) and of the Army Research Laboratory (Grant No. W911NF- 17-1-0072) is gratefully acknowledged.}}
\author{Jin~Dai,~\IEEEmembership{Student~Member,~IEEE,}
        Hai~Lin,~\IEEEmembership{Senior~Member,~IEEE}
\thanks{J. Dai and H. Lin are with the Department of Electrical Engineering, University of Notre Dame, Notre Dame, IN 46556, USA, {\tt\small jdai1@nd.edu, hlin1@nd.edu}.}}
\date{}

\maketitle
\thispagestyle{empty}
\pagestyle{empty}

\begin{abstract}
We investigate the coordination and control problems of distributed discrete event systems that are composed of multiple subsystems subject to potential actuator and/or sensor faults. We model actuator faults as local controllability loss of certain actuator events and sensor faults as observability failure of certain sensor readings, respectively. Starting from automata-theoretic models that characterize behaviors of the subsystems in the presence of faulty actuators and/or sensors, we establish necessary and sufficient conditions for the existence of actuator and sensor fault tolerant supervisors, respectively, and synthesize appropriate local post-fault supervisors to prevent the post-fault subsystems from jeopardizing local safety requirements. Furthermore, we apply an assume-guarantee coordination scheme to the controlled subsystems for both the nominal and faulty subsystems so as to achieve the desired specifications of the system. A multi-robot coordination example is used to illustrate the proposed coordination and control architecture.
\end{abstract}
\begin{IEEEkeywords}
Discrete event systems, supervisor synthesis, sensor and actuator faults, fault tolerant control, coordination.
\end{IEEEkeywords}

\section{Introduction}
\IEEEPARstart{T}{he} ubiquitous deployment of information technology (IT) components has enabled persistent monitoring, coordination and control of large-scale engineering systems with distributed architectures, such as power grids, intelligent transportation systems, cooperative robotic teams and so on. Nevertheless, the application of the heterogeneous IT components has also made the system architectures more sophisticated, rendering them more vulnerable to unpredictable faults that may cause undesired or even catastrhophic consequences. Hence, how to detect and diagnose faults and how to guarantee safe and reliable operation of the engineering systems when faults occur are of great practical importance.

Due to the fact that operation of many engineering systems show strong event-driven features, discrete event system (DES) \cite{cassandras2008introduction} models have become particularly useful for the study of both fault diagnosis and isolation (FDI) and fault tolerant control (FTC) problems \cite{blanke2006diagnosis} over the past decades. Initiated by Sampath et al. \cite{sampath1995diagnosability} in which faults are modeled as unobservable events, many contributions have been made to solving the FDI problem of systems modeled as DESs from different aspects, see, e.g., \cite{contant2006diagnosability, qiu2006decentralized, su2005global, schmidt2013verification, zaytoon2013overview} and the references therein.

Despite the extensive studies on the FDI problem, relatively less work has been done on the FTC problem of DESs. In \cite{lafortune1991tolerable}, a framework is established for synthesizing supervisors for DESs to accomplish both desired and tolerable control objectives. Resilience and fault tolerance of Petri nets are assured in \cite{iordache2004resilience} by using system reconfigurations. Inspired by the adaptive and robust supervisory control techniques \cite{lin1993robust}, respectively, the faults can be addressed using either a passive or an active approach \cite{blanke2006diagnosis}. The {\it passive} FTC is achieved via robust control techniques and a unified controller is designed to ensure that the closed-loop system remains insensitive to certain faults. In contrast, the {\it active} FTC requires a controller that adapts its control policies in case of a fault detection. By following a passive approach, Rohloff \cite{rohloff2005sensor} derives a robust supervisor to assure fault tolerance in which a sensor fault may take place, whereas S\'{a}nchez and Montoya \cite{sanchez2006safe} expand this method by taking safety enforcement into consideration. By assuming some events to be not possible in the presence of a fault, S\"{u}lek and Schmidt \cite{sulek2013computation} present necessary and sufficient conditions for the existence of a fault tolerant supervisor. Wen et al. \cite{wen2008framework} adopt a passive FTC architecture and present the necessary and sufficient condition for the existence of a unique fault tolerant supervisor that can enforce a nominal specification for the non-faulty plant and a tolerable specification for the overall plant. The passive FTC approach is also applied in \cite{karimadini2011fault} for cooperative tasking of multi-agent systems modeled as DESs. To cope with supervisor faults, the reliable supervisory control problem is addressed in \cite{takai2000reliable} and \cite{liu2010reliable} via a passive approach. On the other hand, an active FTC framework of DESs that involves fault detection is presented in \cite{kumar2012framework} to reconfigure the controller so that desired post-fault performance can be met. Darabi et al. \cite{darabi2003control} propose an active FTC method to address faulty sensors by developing a control-switching theory. Paoli et al. \cite{paoli2011active} propose an active FTC framework by safely switching to a post-fault supervisor in response to the online diagnostic information. State-feedback and state-estimate-feedback supervisors are designed in \cite{shu2014fault} to ensure active fault tolerance and safety of a DES. Most of the aforementioned results, however, deal with a centralized monolithic system; whereas the FTC problem of DESs with distributed architectures are not considered.

In this paper, we assume that each subsystem of distributed DESs that are composed of some subsystems
is modeled as a finite automaton and is controlled by a nominal supervisor so that a global specification can be achieved. The fault tolerance property requires that the nominal and fault-pruned subsystems coordinate so as to fulfill the control specification after occurrences of the faults. We propose an active approach for the FTC problem of the distributed DESs subject to possible loss of actuating and sensing capabilities. Specifically, by modeling actuator faults as loss of local controllability of certain actuator events, an active FTC architecture is developed in which the local supervisor is reconfigured to achieve a degraded but safe post-fault performance. Secondly, by characterizing sensor faults as permanent observability failure of certain sensor readings, we develop an automata-theoretic modeling framework of a controlled subsystem in the presence of various faulty sensors, upon which appropriate local supervisors are carried out corresponding to different faulty modes of the subsystem. Furthermore, we allow occurrences of multiple actuator and sensor faults, and we investigate the local FTC problem under multiple faults by introducing a novel model of switching DESs. Finally, by leveraging the idea of compositional verification \cite{cobleigh2003learning}, we present an assume-guarantee post-fault coordination architecture after the synthesis of local post-fault supervisors so as to ensure the accomplishment of the global specification (with possible degradation). Compared to our previous conference publication \cite{dai2015learning} on FTC of multi-agent systems modeled as DESs, we propose different definitions of actuator and sensor faults along with novel synthesis methods of post-fault supervisors in this paper. In addition, more rigorous coordination schemes among post-fault subsystems are developed by accounting for safe operations of each subsystem with faults.

This paper is organized as follows. Section~\ref{sec:preliminaries} reviews the supervisory control theory of DESs modeled as finite automata. The actuator and sensor faults are defined in Section~\ref{sec:formulation} and the fault tolerant coordination and control problem is formulated. We present necessary and sufficient conditions for the existence of an actuator fault tolerant supervisor and guidelines to test them in Section~\ref{sec:actuator}. Section~\ref{sec:sensor} establishes necessary and sufficient conditions for the existence of an sensor fault tolerant supervisor and presents a safe diagnosis and active sensor fault tolerant control architecture. We exploit an assume-guarantee paradigm to achieve post-fault coordination in Section~\ref{sec:discussion}. A multi-robot coordination example is used to illustrate the proposed fault tolerant coordination and control framework in Section~\ref{sec:experiment}. Section~\ref{sec:conclusion} concludes this paper.

\section{Preliminaries}\label{sec:preliminaries}
Let $\Sigma$ be a finite set of {\it events}, and let $2^\Sigma$ and $|\Sigma|$ denote the power set and cardinality of $\Sigma$, respectively. Let $\Sigma^*$ denote the set of all finite-length strings over $\Sigma$ plus the empty string $\epsilon$. For any string $w\in\Sigma^*$, we denote by $|w|$ its length with $|\epsilon|=0$. A language $L$ over $\Sigma$ is a subset $L\subseteq\Sigma^*$. We denote by $L_1-L_2$ the set-theoretic difference of two languages $L_1$ and $L_2$. The {\it prefix-closure} of a language $L$ is a set $\overline{L}:=\{s\in\Sigma^*|(\exists t\in\Sigma^*)[st\in L]\}$. $L$ is said to be {\it prefix-closed} if $\overline{L}=L$. For any string $s\in\Sigma^*$, we denote by $L/s=\{t\in\Sigma^*|st\in L\}$ the set of continuations of $s$ in $L$.

A DES is modeled by a {\it deterministic finite automaton} (DFA) $G=(Q,\Sigma,\delta,q_0,Q_m)$, where $Q$ is the finite set of states, $\Sigma$ is the finite set of events, $\delta: Q\times \Sigma \to Q$ is the (partial) transition function, $q_0\in Q$ is the initial state and $Q_m\subseteq Q$ is the set of marked states. The transition function $\delta$ can be extended to $\delta: Q\times \Sigma^* \to Q$ recursively by: $\delta(q,s\sigma)=\delta(\delta(q,s),\sigma)$, where $q\in Q$, $s\in\Sigma^*$ and $\sigma\in\Sigma$. The language generated by $G$ is defined by $L(G)=\left\{s\in \Sigma^* \vert \delta(q_0,s)!\right\}$, where $\delta(q_0,s)!$ means that $\delta(q_0,s)$ is defined. The language marked by $G$ is $L_m(G)=\left\{s\in \Sigma^* \vert s\in L(G), \delta(q_0,s)\in Q_m\right\}$. We will omit $Q_m$ and write an automaton by $G=(Q,\Sigma, \delta,q_0)$ when marking is not considered. For any $q\in Q$, we denote by $Act_G(q):=\{\sigma\in\Sigma|\delta(q,\sigma)!\}$ the set of events defined at $q$.


For any subset $\Sigma'\subseteq\Sigma$, the {\it natural projection} \cite{cassandras2008introduction} $P_{\Sigma'}$ from $\Sigma^*$ to $\Sigma'^*$ is a mapping $P_{\Sigma'}: \Sigma^*\to \Sigma'^*$  such that: (i) $P_{\Sigma'}(\epsilon)=\epsilon$; (ii) $P_{\Sigma'}(\sigma)=\sigma$ if $\sigma\in\Sigma'$ and $P_{\Sigma'}(\sigma)=\epsilon$ otherwise; (iii) $\forall s\in \Sigma^*, \sigma\in \Sigma$, $P_{\Sigma'}(s\sigma)=P_{\Sigma'}(s)P_{\Sigma'}(\sigma)$. The corresponding inverse projection $P_{\Sigma'}^{-1}:\Sigma'^* \to 2^{\Sigma^*}$ of $P_{\Sigma'}$ is defined as $P_{\Sigma'}^{-1}(t)=\left\{s\in \Sigma^*\vert P_{\Sigma'}(s)=t\right\}$. Functions $P_{\Sigma'}$ and $P_{\Sigma'}^{-1}$ can be extended to a language by applying them to all the strings that belong to the language.

The following notion of ``property satisfaction" relation can then be defined in terms of the natural projection.
\begin{definition}[Satisfaction]\label{satisfaction}
Given a system modeled by a DFA $M=(Q,\Sigma_M,\delta,q_{0,M},Q_{m,M})$ and a property $P$ that is marked by a DFA $P=(Q_P,\Sigma_P, \delta_P,q_{0,P},Q_{m,P})$ with $\Sigma_P \subseteq \Sigma_M$, the system $M$ is said to satisfy $P$, written as $M \models P$, if and only if $\forall t\in L_m(M): P_P(t)\in L_m(P)$, where $P_P$ is the natural projection from $\Sigma_M^*$ to $\Sigma_P^*$.
\end{definition}

\begin{remark}
When $\Sigma_M=\Sigma_P$, Definition~\ref{satisfaction} reduces to the language inclusion $L_m(M)\subseteq L_m(P)$. When both $Q_{m,M}$ and $Q_{m,P}$ are omitted, the satisfaction relation is equivalent to $(\forall t\in L(M))[P_P(t)\in L(P)]$.
\end{remark}

In the Ramadge-Wonham supervisory control theory of DESs modeled as finite automata \cite{ramadge1987supervisory, ramadge1989control}, the event set $\Sigma$ is partitioned into the set of controllable events $\Sigma_c$ and the set of uncontrollable events $\Sigma_{uc}$. When the uncontrolled system is partially observed, $\Sigma$ is also partitioned into the set of observable events $\Sigma_o$ and the set of unobservable events $\Sigma_{uo}$. We associate with $\Sigma_o$ the natural projection $P_o:\Sigma^*\to\Sigma_o^*$. A control decision $\gamma\in 2^\Sigma$ is said to be {\it admissible} if $\Sigma_{uc}\subseteq\gamma$, i.e., uncontrollable events can never be disabled. We define $\Gamma=\{\gamma\in 2^\Sigma|\Sigma_{uc}\subseteq\gamma\}$ as the set of admissible control decisions. Since a supervisor can only make decisions based on its observations, a partial-observation supervisor is a mapping $\mathbb{S}: P_o[L(G)]\to \Gamma$; more specifically, when the plant $G$ generates a string $s$, the supervisor observes $P_o(s)$ and enables events in $\mathbb{S}[P_o(s)]$ accordingly. In practice, the supervisor $\mathbb{S}$ is often implemented as a DFA $S=(Z,\Sigma,\xi,z_0,Z_m)$ such that: (i) for each $z\in Z$ and $\sigma\in\Sigma_{uo}$, $\xi(z,\sigma)!\Rightarrow \xi(z,\sigma)=z$; (ii) for each $s\in L(S)$, $Act_S(\xi(s))=\mathbb{S}[P_o(s)]$. The closed-loop behaviors generated by $G$ under the supervision of $S$ are then given by $L(S\| G)$, where $S\|G$ represents the {\it parallel composition} of two DFAs $S$ and $G$ \cite{cassandras2008introduction,puasuareanu2008learning}.



Given a non-empty and prefix-closed specification language $L=\overline{L}$ over $\Sigma$, $L$ is {\it controllable} (with respect to $G$ and $\Sigma_{uc}$) if $\overline{L}\Sigma_{uc}\cap L(G)\subseteq\overline{L}$; $L$ is {\it observable} (with respect to $G$ and $\Sigma_o$) if $(\forall s, t\in \overline{L}, \sigma\in \Sigma)$ $[P_o(s)=P_o(t)\land s\sigma\in \overline{L}\land t\sigma\in L(G)\Rightarrow t\sigma\in \overline{L}]$. It is well known that there exists a supervisor $S$ such that $L(S\| G)=L$ if and only if $L$ is controllable and observable \cite{ramadge1987supervisory,ramadge1989control,cassandras2008introduction,kumar1995modeling}; in this case, a DFA $S=(Z,\Sigma,\xi,z_0)$ such that $L(S)=L$ suffices to be a satisfactory supervisor. Otherwise, a supervisor $S$ can be synthesized to satisfy $L$ in a {\it maximally permissive} manner, i.e., $L(S\|G)\subseteq L$ and for any other $S'$ such that $L(S'\|G)\subseteq L$, it holds that $L(S\|G)\not\subset L(S'\|G)$ \cite{yin2016synthesis}.

\section{Problem Formulation}\label{sec:formulation}


\subsection{Distributed Discrete Event Systems subject to Faults}

The distributed DES $G$ under consideration is composed of $n$ collaborating subsystems with unique identities, namely $I=\{1,2,\ldots,n\}$. Each subsystem is modeled as an accessible \cite{cassandras2008introduction} DFA $G_i=(Q_i,\Sigma_i,\delta_i,q_{i,0})$ $(i\in I)$. The global event set is defined as $\Sigma=\cup_{i\in I} \Sigma_i$ and we denote by $P_i$ the natural projection from $\Sigma^*$ to $\Sigma_i^*$. For any $\sigma\in\Sigma$, we denote by $In(\sigma)=\{i\in I|\sigma\in\Sigma_i\}$. For each $i\in I$, the local event set $\Sigma_i$ is partitioned into the set of locally controllable events $\Sigma_{i,c}$ and the set of locally uncontrollable events $\Sigma_{i,uc}$. We assume that
\begin{equation}\label{eqn:share}
    (\forall i,j \in I: i\ne j)[\Sigma_{i,uc}\cap\Sigma_{j,c}=\varnothing].
\end{equation}
$\Sigma_i$ is partitioned into the set of locally observable events $\Sigma_{i,o}$ and the set of locally unobservable events $\Sigma_{i,uo}$. Let $P_{i,o}$ denote the natural projection from $\Sigma_i^*$ to $\Sigma_{i,o}^*$. All the subsystems are coordinated via parallel composition, i.e., $G=\|_{i\in I} G_i$.

We are interested in the faults that may interfere with the nominal functionalities of the actuators and/or sensors equipped with the subsystems. In particular, we assume that the nominal supervisor $S_i$ implements the control decisions on the subsystem $G_i$ $(i\in I)$ via $K_i$ local {\it actuators}:
\begin{equation}\label{factuator}
\Sigma_{i,a}=\{\eta_{i,1},\eta_{i,2},\ldots,\eta_{i,K_i}\}\subseteq \Sigma_{i,c}\cap\Sigma_{i,o}.
\end{equation}
More specifically, for any $\eta_{i,m}\in \Sigma_{i,a}$ $(m\in\{1,2,\ldots,K_i\})$, we assume that $In(\eta_{i,m})=\{i\}$; in other words, an actuator can only be enabled or disabled by its own local supervisor. The actuators in $\Sigma_{i,a}$ are assumed to be vulnerable to malfunctions, and we refer to $h_{i,m}$ as the {\it actuator fault} event corresponding to the case in which the supervisor loses the local controllability of the actuator event $\eta_{i,m}$ in $G_i$.

\begin{definition}[Actuator Faults]\label{afault}
For $m\in\{1,2,\ldots,K_i\}$, an actuator fault $h_{i,m}$ occurred in the subsystem $G_i$ $(i\in I)$ indicates that the actuator event $\eta_{i,m}\in \Sigma_{i,a}$ becomes locally uncontrollable for $G_i$. The set of possible actuator fault events in $G_i$ is represented by the set $\Sigma_{i,a}^F=\{h_{i,1},h_{i,2},\ldots,h_{i,K_i}\}$.
\end{definition}

According to Definition~\ref{afault}, an undesired control action may not be prohibited by a local supervisor as a consequence of an actuator fault; this setting is hence consistent with the general understanding of actuator faults \cite{carvalho2018detection}. We assume that an actuator fault is permanent and local controllability of a fault actuator cannot be recovered after the fault.


In addition to actuators, for the purpose of monitoring and controlling the subsystem $G_i$ $(i\in I)$, we denote by
\begin{equation}\label{fsensor}
\Sigma_{i,s}=\{\sigma_{i,1},\sigma_{i,2},\ldots,\sigma_{i,N_i}\}\subseteq \Sigma_{i,o}\cap\Sigma_{i,uc}
\end{equation}
the set of local {\it sensor readings} whose occurrences can be detected by the local supervisor $S_i$. In general, the local observability of the sensor readings in $\Sigma_{i,s}$ is suspicious of loss \cite{sanchez2006safe,carvalho2013robust}; such a situation may correspond to the breakdown of the sensors that monitor and record the occurrences of the event. Other than actuators, it is reasonable to assume a sensor reading to be locally uncontrollable since the local supervisor shall not prevent a sensor reading from being received. For all $k\in\{1,2,\ldots,N_i\}$, we use a {\it sensor fault} event $f_{i,k}$ to capture the circumstance that the sensor reading $\sigma_{i,k}\in \Sigma_{i,s}$ fails to be obtained by $S_i$ .

\begin{definition}[Sensor Faults]\label{sfault}
For $k\in\{1,2,\ldots,N_i\}$, a sensor fault $f_{i,k}$ occurred in the subsystem $G_i$ $(i\in I)$ refers to as the loss of local observability of the sensor reading $\sigma_{i,k}\in \Sigma_{i,s}$. The set of possible sensor fault events in $G_i$ is represented by the set $\Sigma_{i,s}^F=\{f_{i,1},f_{i,2},\ldots,f_{i,N_i}\}$.
\end{definition}

We assume that observability loss of a subsystem's sensor reading
is also permanent. To distinguish a faulty sensor reading from a nominal one, we attach a fault label to the corresponding sensor reading event after it fails to be accessed, resulting in the following set of {\it faulty sensor readings}:
\begin{equation}
    \Sigma^f_{i,s}:=\{\sigma_{i,k}^f|\sigma_{i,k}\in\Sigma_{i,s}\}.
\end{equation}

\subsection{Problem Statement}
Given a non-empty and prefix-closed control specification $L\subseteq L(G)$ for the distributed DES $G$, it is shown in the literature \cite{willner1991supervisory,jiang2000decentralized} that under the assumption \eqref{eqn:share}, a {\it nominal} supervisor $S_i$ $(i\in I)$ can be synthesized for $G_i$ such that $L(S_i\|G_i)=L_i$ and $\|_{i\in I} L_i\subseteq L$. Under the supervision of $S_i$ $(i\in I)$, no undesired behaviors will be generated from the subsystem $G_i$ in the nominal operation. However, undesired strings can arise due to the effective control actions of the nominal supervisor on the faulty subsystem. In this case, a fault tolerant coordination and control is required so as to enforce the accomplishment of the global specification before and after the occurrences of potential faults.

It is also required that safe operation of all the subsystems of $G$  be assured in spite of possible loss of local actuating and/or sensing capabilities. The local safety requirements are captured by a non-empty and prefix-closed safety language $L_i^{safe}\subseteq L(G_i)$ associated with each subsystem $G_i$ $(i\in I)$, which includes all the tolerable behaviors of $G_i$ that should be satisfied (in the sense of Definition~\ref{satisfaction}) in both nominal and fault-pruned operations. Without loss of generality, we assume that $L_i\subseteq L_i^{safe}$ holds for all $i\in I$. Formally, the problem that is addressed in this paper can be stated as follows.

\begin{problem}\label{dcccp}
Consider a distributed DES $G=\|_{i\in I}G_i$ in which $G_i$ $(i\in I)$ is controlled by a nominal supervisor $S_i$ $(i\in I)$ in order to satisfy the local safety $L_i^{safe}$ and the prefix-closed global specification $L$. Suppose that for each $i\in I$, subsystem $G_i$ is equipped with actuators $\Sigma_{i,a}$ and sensor readings $\Sigma_{i,s}$ that may be subject to faults, find a family $S_i^F$ of post-fault supervisors for the subsystem $G_i$ such that:

(i) the local safety requirements shall always be satisfied despite the actuator and sensor faults, i.e., $S_i^F\|G_i^F\models L_i^{safe}$, where $G_i^F$ denotes the subsystem $G_i$ subject to the faults;

(ii) for any subset $I'\in 2^I$ of subsystems whose behaviors are influenced by faults and for any $i\in I'$, $S_i^F$ will steer the faulty subsystems in order to meet the global specification, i.e., $\left(\|_{i\in I-I'} S_i\|G_i\right)\|\left(\|_{j\in I'} S_j^F\|G_j^F\right)\models L$.
\end{problem}

In this paper, objective (i) of Problem~\ref{dcccp} is achieved by {\it local fault tolerant control reconfiguration} of the post-fault subsystem(s) (cf. Sections~\ref{sec:actuator} and~\ref{sec:sensor}). Objective (ii) will be fulfilled by leveraging the idea of {\it assume-guarantee post-fault coordination} (cf. Section~\ref{sec:discussion}), in which the synthesized post-fault supervisors will be further refined if necessary to fulfill the global specification.
\section{An Active Fault Tolerant Control Architecture with Actuator Faults}\label{sec:actuator}
From Definition~\ref{afault}, an actuator fault leads to unexpected loss of (local) controllability of the corresponding actuator event; in other words, the generated language of the controlled subsystem $S_i\|G_i$ may deviate from the presumed local behaviors $L_i$ in a faulty mode, since the local supervisor $S_i$ can no longer disable a faulty actuator event so as to prevent an undesired string from being generated. As a consequence, the presence of the faulty actuator may jeopardize the accomplishment of the global specification $L$. In this section, by assuming that actuator fault events are locally observable, we aim at tackling the actuator fault tolerant control problem with an active approach \cite{blanke2006diagnosis}.

\subsection{Construction of Subsystems subject to an Actuator Fault}
Since $\Sigma_{i,a}\subset \Sigma_{i,o}$ holds both before and after the occurrence of an actuator fault, therefore we can assume that an actuator fault can be detected instantly by means of FDI techniques; i.e., for any $i\in I$ and $m\in\{1,2,\ldots,K_i\}$, $h_{i,m}$ is assumed to be locally observable. The possible successive behaviors of a subsystem after the occurrence of an actuator fault are characterized in terms of the following ``suffix automaton".

\begin{definition}[Suffix Automaton]\label{suffix}
The suffix automaton of an accessible DFA $G=(Q,\Sigma,\delta,q_0)$ following a string $t\in L(G)$ is another DFA $G^{suf}(t)=(Q^{suf},\Sigma,\delta^{suf},q^{suf}_0)$, where $q^{suf}_0=\delta(q_0,t)$, $Q^{suf}=\{q\in Q|(\exists s\in L(G)/t)[\delta(q_0^{suf},s)=q]\},$ and $(\forall q\in Q^{suf},\sigma\in\Sigma)[\delta^{suf}(q,\sigma)=\delta(q,\sigma)\cap Q^{suf}]$.
\end{definition}

It can be verified that the suffix automaton $G^{suf}(t)$ preserves all the possible successive behaviors of the DES $G$ after the string $t$, i.e., $L\left(G^{suf}(t)\right)=L(G)/t$.

To prevent unsafe local behaviors from emerging in a faulty mode, the operation of the nominal supervisor should be disabled and it is necessary to switch from the nominal supervisor to a new post-fault supervisor afterwards, yielding an active approach to the actuator fault tolerance. For convenience of presentation, we assume that an actuator $\eta_{i,m}\in\Sigma_{i,a}$ $(m=1,2,\ldots,K_i)$ becomes locally uncontrollable in the subsystem $G_i$ $(i\in I)$ and the actuator fault event $h_{i,m}$ is detected when a string $t_{i,0}\in L_i$ has been generated during the nominal operation. The uncontrolled post-fault subsystem $G_i$ corresponding to the generation of $t_{i,0}$ and the detection of $h_{i,m}$, written as $G_i^{m,a}(t_{i,0})$, is defined as
\begin{equation}\label{gima}
G_i^{m,a}(t_{i,0})=G_i^{suf}(t_{i,0}),
\end{equation}
where different from $G_i$, the controllability and observability status of the post-fault event set $\Sigma_i^{m,a}$ is given by
\begin{equation}\label{aevent1}
\Sigma_{i,c}^{m,a}=\Sigma_{i,c}-\{\eta_{i,m}\}, \Sigma_{i,uc}^{m,a}=\Sigma_{i,uc}\cup\{\eta_{i,m},h_{i,m}\},
\end{equation}
and
\begin{equation}\label{aevent2}
\Sigma_{i,o}^{m,a}=\Sigma_{i,o}\cup\{h_{i,m}\}, \Sigma_{i,uo}^{m,a}=\Sigma_{i,uo},
\end{equation}
respectively. Note that the actuator fault event is not locally controllable. The following example illustrates the construction of the post-fault model of a given subsystem in the form of a suffix automaton.

\begin{example}
For a subsystem $G_i$ shown in Fig.~\ref{fig:giact}, in which $Q_i=\{q_{i,0},q_{i,1},q_{i,2},q_{i,3}\}$ and $\Sigma_i=\{\eta_{i,1},\eta_{i,2},\eta_{i,3},\eta_{i,4}\}$, we assume that $\Sigma_{i,c}=\Sigma_{i,a}=\Sigma_i$.
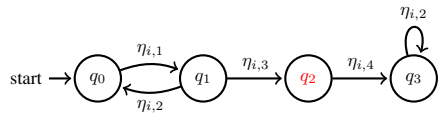
\begin{figure}[h]
	\centering
	\begin{tikzpicture}[shorten >=1pt,node distance=2.0cm,on grid,auto, bend angle=20, thick,scale=0.70, every node/.style={transform shape}]
	\node[state,initial] (s_0)   {$q_0$};
	\node[state] (s_1) [right of = s_0] {$q_1$};
	\node[state] (s_2) [right of = s_1] {$\color{red} q_2$};
    \node[state] (s_3) [right of = s_2] {$q_3$};
	\path[->]
	(s_0) edge [bend left] node [pos=0.5, sloped, above]{$\eta_{i,1}$} (s_1)
	(s_1) edge [bend left] node [pos=0.5, sloped, below]{$\eta_{i,2}$} (s_0)
	(s_1) edge node [pos=0.5, sloped, above]{$\eta_{i,3}$} (s_2)
    (s_2) edge node [pos=0.5, sloped, above]{$\eta_{i,4}$} (s_3)
    (s_3) edge [loop above] node [pos=0.5, sloped, above]{$\eta_{i,2}$} (s_3);
	\end{tikzpicture}
	\caption{The subsystem $G_i$ in the nominal mode.}
    \label{fig:giact}
	\vspace{-4.5mm}
\end{figure}

If the actuator $\eta_{i,2}$ becomes faulty while $G_i$ generates a string $t_{i,0}=\eta_{i,1}\eta_{i,2}\eta_{i,1}\eta_{i,3}$ in the nominal mode, then it follows from \eqref{gima} that the post-fault subsystem $G_i^{2,a}$ is a suffix automaton of $G_i$ and its initial state following should be $q_{i,0}^{suf}=\delta(q_{i,0},t_{i,0})=q_{i,2}$. From Definition~\ref{suffix}, the set of states of $G_i^{2,a}$ is computed as $Q_i^{suf}=\{q_{i,0}^{suf},q_{i,1}^{suf}\}=\{q_{i,2},q_{i,3}\}$. The post-fault subsystem $G_i^{2,a}$ is therefore illustrated in Fig.~\ref{fig:giabac}, with $\eta_{i,2}$ being locally uncontrollable.

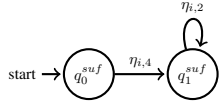
\begin{figure}[h]
	\centering
	\begin{tikzpicture}[shorten >=1pt,node distance=2.3cm,on grid,auto, bend angle=20, thick,scale=0.60, every node/.style={transform shape}]
	\node[state,initial] (s_0)   {$q_0^{suf}$};
	\node[state] (s_1) [right of = s_0] {$q_1^{suf}$};
	\path[->]
	(s_0) edge node [pos=0.5, sloped, above]{$\eta_{i,4}$} (s_1)
    (s_1) edge [loop above] node [pos=0.5, sloped, above]{$\eta_{i,2}$} (s_1);
	\end{tikzpicture}
	\caption{The post-fault subsystem $G_i^{2,a}(t_{i,0})$.}
    \label{fig:giabac}
	\vspace{-7.5mm}
\end{figure}
\end{example}

\subsection{Active Fault Tolerant Control under a Faulty Actuator}
Starting from the post-fault uncontrolled subsystem $G_i^{m,a}$, a set of requirements are posed on the subsystem's controlled behaviors in the faulty mode, resulting in a degraded post-fault specification $L_i^{post}$, which in general is a prefix-closed sublanguage of $L(G_i^{m,a})$. In this paper, we require that the post-fault supervisor $S_i^{m,a}$ be synthesized so that the local safety $L_i^{safe}$ can be maintained. Specifically, if $h_{i,m}$ is detected when a string $t_{i,0}$ is generated, the post-fault specification for $G_i^{m,a}$ is given by:
\begin{equation}\label{ftapost}
L_i^{post}(t_{i,0})=L_i^{safe}/t_{i,0}.
\end{equation}
It can be inferred from Definition~\ref{suffix} and the prefix-closeness of $L_i^{safe}$ that $L_i^{post}(t_{i,0})$ is a prefix-closed sublanguage of $L(G_i^{m,a}(t_{i,0}))$. Considering the post-fault specification $L_i^{post}(t_{i,0})$, the following notion of actuator fault tolerance is presented to justify $G_i^{m,a}$'s capability of assuring local safety.

\begin{definition}[Actuator Fault Tolerance]\label{aftolerance}
Language $L\left(G_i^{m,a}(t_{i,0})\right)$ $(i\in I)$ is said to be {\it actuator fault tolerant} with respect to the faulty actuator $\eta_{i,m}\in\Sigma_{i,a}$ and the string $t_{i,0}\in L_i$ if there exists a non-empty sublanguage of $L_i^{post}(t_{i,0})$ that is controllable with respect to $G_i^{m,a}(t_{i,0})$ and $\Sigma_{i,uc}^{m,a}$ and observable with respect to $G_i^{m,a}(t_{i,0})$ and $\Sigma_{i,o}^{m,a}$.
\end{definition}

\begin{figure}[h]
\begin{center}
    \centerline{\includegraphics[width=0.60\textwidth]{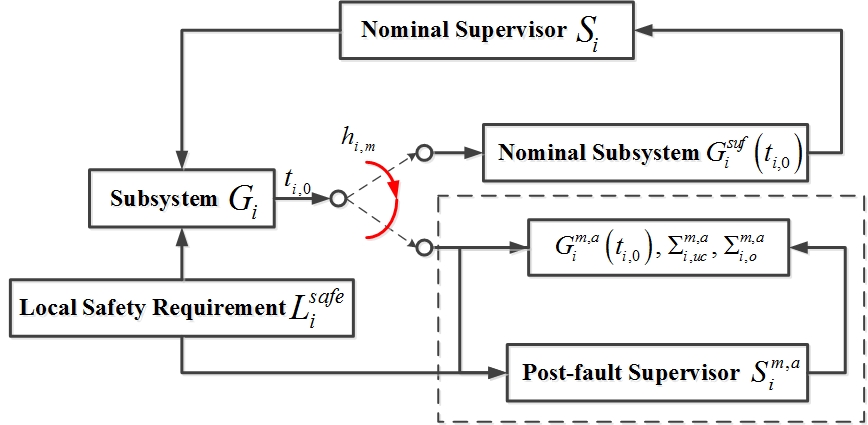}}
    \caption{An actuator fault tolerant control architecture of $G_i$.}
    \label{fta}
  \end{center}
 \vspace{-8.5mm}
\end{figure}

A general architecture is depicted in Fig.~\ref{fta} to achieve active fault tolerant control of the subsystem $G_i$ $(i\in I)$ in the presence of an faulty actuator $\eta_{i,m}$. In the nominal mode, no actuator fault takes place and the control loop is closed on the nominal supervisor $S_i$ that enables appropriate locally controllable events based on the observation $P_{i,o}(t_{i,0})$ when the subsystem generates a string $t_{i,0}$. Once the actuator $\eta_{i,m}$ becomes faulty, the actuator fault event $h_{i,m}$ is generated to interrupt the nominal operation of $S_i$ to prevent any unsafe behaviors from emerging in $G_i$. In this case, we test the actuator fault tolerance of the post-fault subsystem $G_i^{m,a}(t_{i,0})$ and if the actuator fault tolerance holds , there must exist a post-fault supervisor $S_i^{m,a}$ that steers $G_i^{m,a}$ to fulfill the post-fault specification $L_i^{post}(t_{i,0})$.

We now study the procedure of testing the actuator fault tolerance property after an actuator fault. Thanks to the standard $\inf C(\cdot)$ operator, which computes the infimal prefix-closed controllable sublanguage of a given language with respect to another given language and a given set of uncontrollable events \cite{cassandras2008introduction}, we present the following theorem that states the necessary and sufficient conditions of the actuator fault tolerance of $L\left(G_i^{m,a}(t_{i,0})\right)$.

\begin{theorem}\label{atolerance}
Language $L\left(G_i^{m,a}(t_{i,0})\right)$ is actuator fault tolerant with respect to $\eta_{i,m}$ and $t_{i,0}$ if and only if language $\inf C_i^{m,a}(\{\epsilon\})$, computed with respect to $L\left(G_i^{m,a}(t_{i,0})\right)$ and $\Sigma_{i,uc}^{m,a}$, satisfies that $\inf C_i^{m,a}(\{\epsilon\})\subseteq L_i^{post}(t_{i,0})$.
\end{theorem}

\begin{IEEEproof}
By construction, $\inf C_i^{m,a}(\{\epsilon\})$ computed with respect to $L\left(G_i^{m,a}(t_{i,0})\right)$ contains all the shortest continuations of $t_{i,0}$ in $L\left(G_i^{m,a}(t_{i,0})\right)$ after the detection of $h_{i,m}$, in which the post-fault subsystem can be controlled, i.e., all the possible evolutions of $G_i^{m,a}$ after disabling all the local events that can be feasibly disabled. Suppose that there exists a string $s_{i,0}\in\inf C_i^{m,a}(\{\epsilon\})$ but $s_{i,0}\not\in L_i^{post}(t_{i,0})$, this means that there is no way to prevent the string $t_{i,0}s_{i,0}\in L(G_i)-L_i^{safe}$ from emerging in $G_i$, which clearly leads to a violation of the actuator fault tolerance of $G_i^{m,a}$.

Conversely, suppose that the actuator fault tolerance of $L\left(G_i^{m,a}(t_{i,0})\right)$ fails to be satisfied with respect to some $\eta_{i,m}$ and $t_{i,0}$, i.e., there does not exist a post-fault supervisor $S_i^{m,a}$ that can steer $G_i^{m,a}$ to satisfy $L_i^{post}(t_{i,0})$. In this case, there always exists at least one string $s_{i,0}\in L\left(G_i^{m,a}\right)-L_i^{post}(t_{i,0})$ that is formed by uncontrollable events in $\Sigma_{i,uc}^{m,a}$ (otherwise the occurrence of $s_{i,0}$ can be prohibited {\it a priori} by simply disabling any locally controllable event in the string). Since $\inf C_i^{m,a}(\{\epsilon\})$ enumerates all the concatenations of locally uncontrollable events feasible in $L(G_i^{m,a}(t_{i,0}))$, it holds that $s_{i,0}\in\inf C_i^{m,a}(\{\epsilon\})$; therefore $\inf C_i^{m,a}(\{\epsilon\})\not\subseteq L_i^{post}(t_{i,0})$ is satisfied, which leads to a contradiction.
\end{IEEEproof}

Theorem~\ref{atolerance} presents the necessary and sufficient conditions for the existence of a post-fault supervisor that ensures local safety. Once an actuator $\eta_{i,m}$ becomes faulty, a local supervisor that achieves $\inf C_i^{m,a}(\{\epsilon\})$ suffices to be a post-fault supervisor; however, it may not yield a satisfactory solution as $\inf C_i^{m,a}(\{\epsilon\})$ may be too restrictive. To efficiently compute a satisfactory $S_i^{m,a}$, we associate the the subsystem $G_i$ $(i\in I)$ with the following bank of {\it safety supervisors}:
\begin{equation}
\mathbb{S}_i^a=\{S_i^m| m\in\{1,2,\ldots,K_i\}, L(S_i^m\|G_i)\subseteq L_i^{safe}\},
\end{equation}
where each safety supervisor $S_i^m\in\mathbb{S}_i$ can be synthesized offline such that $L_i^m=L(S_i^m\|G_i)\subseteq L_i^{safe}$ is satisfied in a maximally permissive manner, under the assumption that $\eta_{i,m}$ is locally uncontrollable. In the case that $t_{i,0} \in  L_i^m$ for the string $t_{i,0}$ before the detection of $h_{i,m}$, $S_i^{m,a}$ can be implemented offline and is given by the following suffix automaton of $S_i^m$ after the generation of $t_{i,0}$, i.e.,
\begin{equation}\label{sima}
    S_i^{m,a}=S_i^{m^{suf}}(t_{i,0}).
\end{equation}
It then follows from \eqref{sima} that
\begin{equation*}
\begin{split}
L\left(S_i^{m,a}\|G_i^{m,a}\right)&=L(S_i^m\|G_i)/t_{i,0}\\
&\subseteq L_i^{safe}/t_{i,0}=L_i^{post}(t_{i,0}).
\end{split}
\end{equation*}
Therefore, as long as $G_i^{m,a}(t_{i,0})$ is actuator fault tolerant, $S_i^{m,a}$ \eqref{sima} can be extracted from $S_i^m\in\mathbb{S}_i^m$ and is sufficient to fulfill the post-fault specification.

On the other hand, if $t_{i,0}\not\in L_i^m$, the safety supervisor $S_i^m$ cannot be applied directly to synthesize $S_i^{m,a}$. In this case, the computational complexity of synthesizing an offline post-fault supervisor is exponential in the number of states in $G_i^{m,a}$; therefore, we adopt online control techniques for the synthesis of $S_i^{m,a}$ (see, e.g., \cite{heymann1994line,hadj1996centralized}), which generally possess polynomial complexity
at each locally observable event along a trajectory in $L_i^{post}$.

\subsection{Fault Tolerant Control with Multiple Actuator Faults}
In the previous subsection, we discussed local fault tolerant control under the assumption that there are no consecutive occurrences of multiple actuator faults. However, multiple faults may occur in many practical engineering systems. Therefore, it is necessary to extend the proposed active fault tolerant control architecture to address multiple actuator faults. In the presence of multiple faults, the subsystem shall then switch from either the nominal mode to a faulty mode, or one faulty mode to another. We introduce the following set of {\it mode-switching} events to handle multiple faults in $G_i$,
\begin{equation}\label{sigmaswa}
\begin{split}
\Sigma_{i,a}^{sw}=\{h_i^{m_1,m_2}|& m_1,m_2\in\{0,1,2,\ldots,K_i\},\\
& m_1\ne m_2,m_2\ne 0\},
\end{split}
\end{equation}
where we denote by the $0$-th mode the nominal mode. The mode-switching event $h_i^{m_1,m_2}$ is generated when $\eta_{i,m_2}$ becomes faulty in the $m_1$-th faulty mode. For convenience of presentation, we write $h_i^{0,m}=h_{i,m}$, indicating the switch from the nominal mode to the $m$-th faulty mode. It is reasonable to assume $m_2\ne 0$ in \eqref{sigmaswa} since the subsystem cannot return to the nominal mode from a faulty one. Similar to the actuator fault events, all the mode-switching events in $\Sigma_{i,a}^{sw}$ are assumed to be locally observable.

We study the case in which actuators $\eta_{i,1}, \eta_{i,2},\ldots,\eta_{i,K_i}$ of the subsystem $G_i$ may become faulty. When there is no fault, the subsystem stays in the nominal mode, the fault tolerant control is inactive. After some faults occur and the subsystem enters a faulty mode, the active fault tolerant control framework depicted in Fig.~\ref{fta} is inherited here to resolve the impacts of the faults. Without loss of generality, we still assume that the first actuator fault occurs when a string $t_{i,0}$ is generated in $G_i$. Thus the uncontrolled post-fault subsystem model of $G_i$ in the presence of multiple faults is given by
\begin{equation}\label{gifa}
    G_i^{F,a}(t_{i,0})=G_i^{suf}(t_{i,0}).
\end{equation}
Furthermore, we consider that the actuator faults may occur in {\it arbitrary} orders. Toward this regards, we define the controllable and observable events in the presence of $K_i$ possible faulty actuators as follows, respectively:
\begin{equation}\label{aeventm}
\begin{split}
&\Sigma_{i,c}^{F,a}=\Sigma_{i,c}-\cup_{m=1}^{K_i}\{\eta_{i,m}\}=\Sigma_{i,c}-\Sigma_{i,a}, \\
&\Sigma_{i,uc}^{F,a}=\Sigma_{i,uc}\cup\Sigma_{i,a}\cup\Sigma_{i,a}^{sw};\\
&\Sigma_{i,o}^{F,a}=\Sigma_{i,o}\cup\Sigma_{i,a}^{sw}, \Sigma_{i,uo}^{F,a}=\Sigma_{i,uo}.
\end{split}
\end{equation}
The local control objective of the post-fault supervisor for the subsystem $G_i$ subject to multiple actuator faults is to guarantee the accomplishment of the local safety specification $L_i^{safe}$. Therefore, the post-fault specification is obtained as
\begin{equation}\label{ftampost}
\begin{split}
L_i^{post}(t_{i,0})=L_i^{safe}/t_{i,0}.
\end{split}
\end{equation}


Since $h_i^{m_1,m_2}\in \Sigma_{i,a}^{sw}$ is locally observable, we can disable the operation of the nominal supervisor $S_i$ when the first actuator $\eta_{i,m}$ fails to be locally controllable and $h_i^{0,m}$ is detected for some $m\in\{1,2,\ldots,K_i\}$. Following the detection of the mode-switching event, we require that the post-fault supervisor $S_i^{F,a}$ be synthesized in order to ensure the post-fault specification $L_i^{post}(t_{i,0})$ in the presence of arbitrary switch among various faulty modes. The solvability of the fault tolerant control problem under multiple actuator faults is given below.
%
%

\begin{theorem}\label{multiact}
There exists a post-fault supervisors $S_i^{F,a}$ that can steer the faulty subsystem $G_i^{F,a}$ $(i\in I)$ to satisfy $L_i^{post}(t_{i,0})$ \eqref{ftampost} in the presence of faulty actuators in $\Sigma_{i,a}$ if and only if language $\inf C_i^{F,a}(\{\epsilon\})$, computed with respect to $L(G_i^{F,a}(t_{i,0}))$ and $\Sigma_{i,uc}^{F,a}$, satisfies that $\inf C_i^{F,a}(\{\epsilon\})\subseteq L_i^{post}(t_{i,0})$.
\end{theorem}

\begin{IEEEproof}
The proof is similar to that of Theorem~\ref{atolerance} and is omitted.
\end{IEEEproof}

To implement the post-fault supervisor in the multi-fault case, a safety supervisor $S_i^a$ can be synthesized offline to accomplish $L_i^{safe}$ in a maximally permissive manner on condition that all actuator events in $\Sigma_{i,a}$ are locally uncontrollable. Thanks to the safety supervisor $S_i^a$, if the string $t_{i,0}\in L_i$ before the detection of the first mode-switching event satisfies that $t_{i,0}\in L(S_i^a\|G_i)$, then similar to the single-fault case \eqref{sima}, a satisfactory post-fault supervisor $S_i^{F,a}$ can be implemented offline as
\begin{equation}
    S_i^{F,a}=S_i^{a^{suf}}(t_{i,0}).
\end{equation}
Otherwise, $S_i^{F,a}$ shall be implemented via online control techniques \cite{heymann1994line,hadj1996centralized}.

We use $S_i^F\|G_i^F$ to denote the overall closed-loop model of the subsystem $G_i$ $(i\in I)$ with potential faulty actuators in $\Sigma_{i,a}$ both prior to and after the detection of the first mode-switching events $h_i^{0,m}\in\Sigma_{i,a}^{sw}$ for some $m$. The following theorem states that the active fault tolerant architecture proposed in this section will steer the subsystem $G_i$ to ensure local safety.

\begin{theorem}\label{thm:fta}
For the subsystem $G_i$ $(i\in I)$ that is equipped with local actuators in $\Sigma_{i,a}$ whose local controllability may fail in arbitrary orders, the switching among the nominal supervisor $S_i$ and the post-fault supervisors $S_i^{F,a}$ ensures the local safety specification $L_i^{safe}$, i.e., $S_i^F\|G_i^F\models L^{safe}_i$.
\end{theorem}
\begin{IEEEproof}
With slightly abusing the notations, we use $L\left(S_i^F\|G_i^F\right)$ to denote the behaviors of the closed-loop subsystem $S_i^F\|G_i^F$. Since the post-fault supervisor $S_i^{F,a}$ is synthesized to ensure the post-fault specification $L_i^{post}$ regardless of the order of the actuator faults, therefore $L\left(S_i^F\|G_i^F\right)$ should contain the strings that are represented in the form of the concatenation of strings from $L(S_i\|G_i)$ and strings from $L\left(S_i^{F,a}\|G_i^{F,a}\right)$. Formally, we can write that
\begin{equation*}
\begin{split}
L\left(S_i^F\|G_i^F\right)=&\left\{\overline{t_{i,0}h_i^{0,m_1}t_{i,1}h_i^{m_1,m_2}t_{i,2}\cdots}\right. \\
&\left.\overline{t_{i,K_i-1}h_i^{m_{K_i-1},m_{K_i}}t_{i,K_i}}\right\},
\end{split}
\end{equation*}
where $m_1,m_2,\ldots,m_{K_i}$ is an enumeration of $\{1,2,\ldots,K_i\}$ and $\overline{t_{i,0}}\subseteq L_i\subseteq L_i^{safe}$. Furthermore, according to \eqref{ftampost}, it holds that
\begin{equation*}
\begin{split}
h_i^{0,m_1}t_{i,1}h_i^{m_1,m_2}t_{i,2}\cdots \\
t_{i,m_{K_i-1}}h_i^{m_{K_i-1},m_{K_i}}t_{i,m_{K_i}} &\subseteq L_i^{post}\\
&=L_i^{safe}/t_{i,0}.
\end{split}
\end{equation*}
Let $P_i^{F,a}$ be the natural projection from $\left[\Sigma_i\cup\Sigma_{i,a}^{sw}\right]^*$ to $\Sigma_i^*$. We can then write that
\begin{equation}\label{20}
P_i^{F,a}\left[L\left(S_i^{F,a}\|G_i^{F,a}\right)\right]=\overline{t_{i,0}t_{i,1}\cdots t_{i,K_i}}\subseteq L_i^{safe}
\end{equation}
can always be satisfied. By Definition~\ref{satisfaction}, \eqref{20} is equivalent to $S_i^{F}\|G_i^{F}\models L_i^{safe}$, which completes the proof.
\end{IEEEproof}

\section{Safe Supervisory Control with Sensor Faults}\label{sec:sensor}
This section is concerned with the synthesis of the local post-fault supervisor for a subsystem whose nominal operation suffers from sensor faults. The proposed fault tolerant supervisory control scheme includes two major ingredients: (i) construction of an automaton model of the subsystem in the presence of local sensor faults; (ii) synthesis of the corresponding post-fault supervisor(s) after the sensor faults.

\subsection{Modeling Sensor Faults in Discrete Event Systems}
Let $G_i^0=S_i\|G_i$ denote the nominal mode of the subsystem $G_i$ $(i\in I)$ under control. Since undesired behaviors may arise after occurrences of potential faults as a consequence of executing nominal supervision commands on the fault-pruned subsystem, we aim at exploring the behaviors generated by $G_i^0$ when one sensor reading $\sigma_{i,k}\in\Sigma_{i,s}$ becomes faulty. As defined in \eqref{fsensor}, the suspicious sensor readings in $\Sigma^f_{i,s}$ introduce $N_i$ faulty modes to the subsystem $G_i$ in addition to the nominal mode. For $k\in\{1,2,\ldots,N_i\}$, we assume that the $k$-th faulty mode of $G_i$ is modeled as
\begin{equation}
G_{i,k}=(Q_{i,k},\Sigma_{i,k},\delta_{i,k}),\quad k=1,2,\ldots,N_i,
\end{equation}
where $Q_{i,k}=\{q_{i,k,l}|q_{i,l}\in Q_i\}$ is a copy of $Q_i$, $\Sigma_{i,k}=\Sigma_i\cup\{f_{i,k},\sigma_{i,k}^f\}$, and the transition function $\delta_{i,k}$ is defined as follows: for any $q_{i,k,l}\in Q_{i,k}$ and $\sigma\in\Sigma_{i,k}$,
\begin{equation*}
    \delta_{i,k}(q_{i,k,l},\sigma)=
    \begin{cases}
    q_{i,k,l'}, \mbox{ if } \sigma\in\Sigma_i\land\delta_i(q_{i,l},\sigma)=q_{i,l'};\\
    q_{i,k,l'}, \mbox{ if } \sigma=\sigma^f_{i,k}\land\delta_i(q_{i,l},\sigma_{i,k})=q_{i,l'}.
    \end{cases}
\end{equation*}
Note that the initial state $q_{i,k,0}$ of $G_{i,k}$ is not specified at this point since it depends on at which state of $G_i$ the sensor fault $f_{i,k}$ occurs. Furthermore, we define the transition $\delta_{i,k}(q_{i,l},\sigma_{i,k}^f)$ in addition to $\delta_{i,k}(q_{i,l},\sigma_{i,k})$ in the $k$-th faulty mode. For convenience of presentation, we also use $f_{i,k}$ to denote the associated mapping
\begin{equation*}
f_{i,k}: Q_i\to Q_{i,k},\quad k=1,2,\ldots,N_i,
\end{equation*}
i.e., the sensor fault $f_{i,k}$ indicates the transition from the nominal mode $G_i$ to the $k$-th faulty mode $G_{i,k}$. In other words, $f_{i,k}(q_{i,l})=q_{i,k,l}$ indicates that $f_{i,k}$ occurs when $G_i$ evolves to $q_{i,l}$ and the initial state of $G_{i,k}$ turns to be $q_{i,k,l}$. Since we pose no assumption on when the sensor fault may occur, the {\it unified} model of $G_i$ subject to one suspicious sensor reading $\sigma_{i,k}$ can then be constructed as
\begin{equation}\label{gikmodel}
G_i^k=(Q_i^k,\Sigma_i^k,\delta_i^k,q_{i,0}^k),
\end{equation}
where $Q_i^k=Q_i\cup Q_{i,k}$, $\Sigma_i^k=\Sigma_{i,k}$, $q_{i,0}^k=q_{i,0}$, with the transition function $\delta_i^k=\delta_i\cup\delta_{i,k}\cup\{(q_{i,l},f_{i,k},q_{i,k,l})\vert q_{i,l}\in Q_i\}$.

In addition to $G_i^k$, we proceed to the construction of $S_i^k$, which is the counterpart of $S_i$ in the $k$-th faulty mode. For convenience of presentation, we assume that the nominal supervisor $S_i$ can be realized as the following DFA:
\begin{equation}\label{si}
S_i=(X_i,\Sigma_i,\xi_i^0,x_{i,0}).
\end{equation}
Similar to $G_{i,k}$, we also use $S_{i,k}$ to denote the potential behaviors of $S_i$ in the $k$-th faulty mode; more specifically, $S_{i,k}$ is given by the following automaton without a specified initial state:
\begin{equation*}
    S_{i,k}=(X_{i,k},\Sigma_{i,k},\xi_{i,k}),\quad k=1,2,\ldots,N_i,
\end{equation*}
where $X_{i,k}=\{x_{i,k,l}|x_{i,l}\in X_i\}$, and for any $x_{i,k,l}\in X_{i,k}$ and $\sigma\in \Sigma_{i,k}$, $\xi_{i,k}(x_{i,k,l},\sigma)$ is formally defined as
\begin{equation*}
    \xi_{i,k}(x_{i,k,l},\sigma)=
    \begin{cases}
    x_{i,k,l'}, &\mbox{if } \xi_i(x_{i,l},\sigma)=x_{i,l'}; \\
    x_{i,k,l}, &\mbox{if } [\sigma=\sigma_{i,k}^f \land \xi_i(x_{i,l},\sigma_{i,k})!]\lor \\
               &[\sigma\in\Sigma_{i,uc}\cup\{\sigma_{i,k}^f\} \land \neg\xi_i(x_{i,l},\sigma)!].
    \end{cases}
\end{equation*}
In other words, we make sure that uncontrollable events shall never be disabled regardless of possible sensor faults by adding self-loops that are labeled by uncontrollable events and faulty sensor readings to all the states in the faulty mode. Therefore, the {\it unified} model of $S_i$ in the presence of the unobservalbe sensor reading $\sigma_{i,k}^f$ is obtained as
\begin{equation}\label{sikmodel}
S_i^k=(X_i^k,\Sigma_i^k,\xi_i^k,x_{i,0}^k),
\end{equation}
where $X_i^k=X_i\cup X_{i,k}$, $x_{i,0}^k=x_{i,0}$ and $\xi_i^k=\xi_i\cup\xi_{i,k}\cup\{(x_{i,l},f_{i,k},x_{i,k,l})\vert x_{i,l}\in X_i\}$. Finally, by leveraging \eqref{gikmodel} and \eqref{sikmodel}, the closed-loop model of the controlled subsystem $G_i^0$ in the presence of a faulty sensor reading $\sigma_{i,k}^f$ is computed as
\begin{equation}\label{giks}
    G_i^{k,s}=S_i^k\| G_i^k:=(Q_i^{k,s},\Sigma_i^{k,s},\delta_i^{k,s},q_{i,0}^{k,s}),
\end{equation}
where the set of post-fault local events is given by $\Sigma_i^{k,s}=\Sigma_i\cup\{\sigma_{i,k}^f,f_{i,k}\}$ and can be further partitioned as
\begin{equation}\label{sensorevent}
\begin{split}
&\Sigma_{i,c}^{k,s}=\Sigma_{i,c},\Sigma_{i,uc}^{k,s}=\Sigma_{i,uc}\cup\{\sigma_{i,k}^f,f_{i,k}\};\\
&\Sigma_{i,o}^{k,s}=\Sigma_{i,o},\Sigma_{i,uo}^{k,s}=\Sigma_{i,uo}\cup\{\sigma_{i,k}^f,f_{i,k}\}.
\end{split}
\end{equation}
Note that each step \eqref{gikmodel}---\eqref{giks} of the construction procedure of $G_i^{k,s}$ requires no prior knowledge of the state at which the sensor fault event occurs and hence can be computed {\it offline}.

We use the following example to demonstrate the construction procedure of $G_i^{k,s}$ .
\begin{example}
(Rephrased from \cite{carvalho2018detection}, Example 4) We consider a subsystem $G_i$ $(i\in I)$ whose DFA representation is depicted in Fig.~\ref{fig:gi}, where $\Sigma_i=\Sigma_{i,o}=\{a,b,c\}$, $\Sigma_{i,c}=\{a,c\}$. Let $\Sigma_{i,s}=\{b\}$ be the set of suspicious sensor readings. The safety language is given by $L_i^{safe}=\overline{ab+ac}$, i.e., the state $5$ (marked with double circles) is an unsafe state. A nominal supervisor $S_i$ can be synthesized to ensure $L_i^{safe}$ and is shown in Fig.~\ref{fig:si}.
\begin{figure}[h]
	\centering
	\begin{tikzpicture}[shorten >=1pt,node distance=2.0cm,on grid,auto, bend angle=20, thick,scale=0.75, every node/.style={transform shape}]
	\node[state,initial] (s_1)   {$1$};
	\node[state] (s_2) [right of = s_1] {$2$};
	\node[state] (s_4) [right of = s_2] {$4$};
	\node[state,accepting] (s_5) [right of = s_4] {$5$};
    \node[state] (s_3) [below right of = s_2] {$3$};
	\path[->]
	(s_1) edge node [pos=0.5, sloped, above]{$a$} (s_2)
	(s_2) edge node [pos=0.5, sloped, above]{$b$} (s_4)
    (s_4) edge node [pos=0.5, sloped, above]{$c$} (s_5)
    (s_2) edge node [pos=0.5, sloped, above]{$c$} (s_3);

	\end{tikzpicture}
	\caption{The DFA model of $G_i$.}
    \label{fig:gi}
	\vspace{-5mm}
\end{figure}
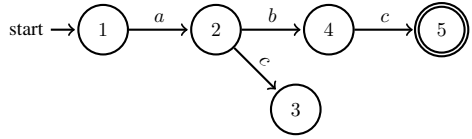

\begin{figure}[h]
	\centering
	\begin{tikzpicture}[shorten >=1pt,node distance=2.0cm,on grid,auto, bend angle=20, thick,scale=0.75, every node/.style={transform shape}]
	\node[state,initial] (s_1)   {$1$};
	\node[state] (s_2) [right of = s_1] {$2$};
	\node[state] (s_4) [right of = s_2] {$4$};
    \node[state] (s_3) [below right of = s_2] {$3$};
	\path[->]
	(s_1) edge node [pos=0.5, sloped, above]{$a$} (s_2)
	(s_2) edge node [pos=0.5, sloped, above]{$b$} (s_4)
    (s_2) edge node [pos=0.5, sloped, above]{$c$} (s_3);

	\end{tikzpicture}
	\caption{The DFA model of $S_i$.}
    \label{fig:si}
	\vspace{-4mm}
\end{figure}
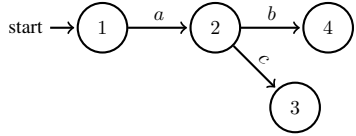

Next, following the construction procedures \eqref{gikmodel} and \eqref{sikmodel}, we can compute $G_i^k$ and $S_i^k$ accordingly that are shown in Fig.~\ref{fig:gik} and Fig.~\ref{fig:sik}, respectively.
\begin{figure}[h]
	\centering
	\begin{tikzpicture}[shorten >=1pt,node distance=2.0cm,on grid,auto, bend angle=20, thick,scale=0.75, every node/.style={transform shape}]
	\node[state,initial] (s_1)   {$1$};
	\node[state] (s_2) [right of = s_1] {$2$};
	\node[state] (s_4) [right of = s_2] {$4$};
	\node[state,accepting] (s_5) [right of = s_4] {$5$};
    \node[state] (s_3) [below right of = s_2] {$3$};
    \node[state] (s_13) [below of = s_3] {$3'$};
    \node[state] (s_12) [below left of = s_13] {$2'$};
    \node[state] (s_11) [left of = s_12] {$1'$};
	\node[state] (s_14) [right of = s_12] {$4'$};
	\node[state,accepting] (s_15) [right of = s_14] {$5'$};

	\path[->]
	(s_1) edge node [pos=0.5, sloped, above]{$a$} (s_2)
	(s_2) edge node [pos=0.5, sloped, above]{$b$} (s_4)
    (s_4) edge node [pos=0.5, sloped, above]{$c$} (s_5)
    (s_2) edge node [pos=0.5, sloped, above]{$c$} (s_3)
    (s_11) edge node [pos=0.5, sloped, above]{$a$} (s_12)
    (s_14) edge node [pos=0.5, sloped, above]{$c$} (s_15)
    (s_12) edge node [pos=0.5, sloped, above]{$c$} (s_13)
    (s_12) edge [bend left] node [pos=0.5, sloped, above]{$b$} (s_14)
    (s_12) edge [bend right] node [pos=0.5, sloped, below]{$b^f$} (s_14)
    (s_1) edge node [pos=0.5, sloped, above]{$f_{b}$} (s_11)
    (s_2) edge node [pos=0.5, sloped, above]{$f_{b}$} (s_12)
    (s_3) edge node [pos=0.5, sloped, below]{$f_{b}$} (s_13)
    (s_4) edge node [pos=0.5, sloped, above]{$f_{b}$} (s_14)
    (s_5) edge node [pos=0.5, sloped, above]{$f_{b}$} (s_15);

	\end{tikzpicture}
	\caption{The DFA model of $G_i^k$.}
    \label{fig:gik}
	\vspace{-4mm}
\end{figure}
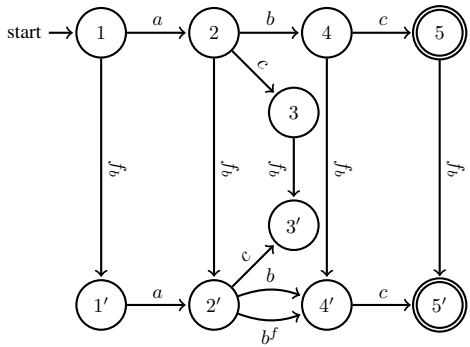

\begin{figure}[h]
	\centering
	\begin{tikzpicture}[shorten >=1pt,node distance=2.5cm,on grid,auto, bend angle=20, thick,scale=0.75, every node/.style={transform shape}]
	\node[state,initial] (s_1)   {$1$};
	\node[state] (s_2) [right of = s_1] {$2$};
	\node[state] (s_4) [right of = s_2] {$4$};
    \node[state] (s_3) [below right of = s_2] {$3$};
    \node[state] (s_13) [below of = s_3] {$3'$};
    \node[state] (s_12) [below left of = s_13] {$2'$};
    \node[state] (s_11) [left of = s_12] {$1'$};
	\node[state] (s_14) [right of = s_12] {$4'$};

	\path[->]
	(s_1) edge node [pos=0.5, sloped, above]{$a$} (s_2)
	(s_2) edge node [pos=0.5, sloped, above]{$b$} (s_4)
    (s_2) edge node [pos=0.5, sloped, above]{$c$} (s_3)
    (s_11) edge node [pos=0.5, sloped, above]{$a$} (s_12)
    (s_12) edge node [pos=0.5, sloped, above]{$c$} (s_13)
    (s_12) edge node [pos=0.5, sloped, above]{$b$} (s_14)
    (s_11) edge [loop below] node[pos=0.5, sloped, below]{$b,b^f$} (s_11)
    (s_12) edge [loop below] node[pos=0.5, sloped, below]{$b^f$} (s_12)
    (s_14) edge [loop below] node[pos=0.5, sloped, below]{$b,b^f$} (s_14)
    (s_1) edge node [pos=0.5, sloped, above]{$f_{b}$} (s_11)
    (s_2) edge node [pos=0.5, sloped, above]{$f_{b}$} (s_12)
    (s_3) edge node [pos=0.5, sloped, below]{$f_{b}$} (s_13)
    (s_4) edge node [pos=0.5, sloped, above]{$f_{b}$} (s_14);

	\end{tikzpicture}
	\caption{The DFA model of $S_i^k$.}
    \label{fig:sik}
	\vspace{-4mm}
\end{figure}
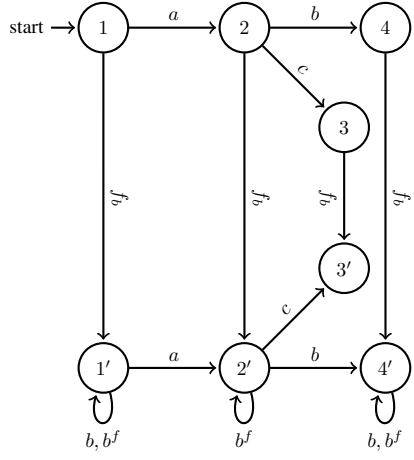

Finally, we can obtain $G_i^{k,s}$ by composing $G_i^k$ and $S_i^k$. As shown in Fig.~\ref{fig:giks}, whenever the sensor fault $f_b$ takes place before the occurrence of the sensor reading $b$, the closed-loop system may still allow the occurrence of unsafe behaviors (in this example, we can see that the string $af_bb_fc\not\models L_i^{safe}$).
\begin{figure}[h]
	\centering
	\begin{tikzpicture}[shorten >=1pt,node distance=2.7cm,on grid,auto, bend angle=20, thick,scale=0.75, every node/.style={transform shape}]
	\node[state,initial] (s_1)   {$(1,1)$};
	\node[state] (s_2) [right of = s_1] {$(2,2)$};
	\node[state] (s_4) [right of = s_2] {$(4,4)$};
    \node[state] (s_3) [below right of = s_2] {$(3,3)$};
    \node[state] (s_13) [below of = s_3] {$(3',3')$};
    \node[state] (s_12) [below left of = s_13] {$(2',2')$};
    \node[state] (s_11) [left of = s_12] {$(1',1')$};
	\node[state] (s_14) [right of = s_12] {$(4',4')$};
	\node[state] (s_22) [below right of = s_12] {$(2',4')$};
	\node[state] (s_23) [right of = s_22] {$(3',5')$};

	\path[->]
	(s_1) edge node [pos=0.5, sloped, above]{$a$} (s_2)
	(s_2) edge node [pos=0.5, sloped, above]{$b$} (s_4)
    (s_2) edge node [pos=0.5, sloped, above]{$c$} (s_3)
    (s_11) edge node [pos=0.5, sloped, above]{$a$} (s_12)
    (s_12) edge node [pos=0.5, sloped, above]{$c$} (s_13)
    (s_12) edge node [pos=0.5, sloped, above]{$b$} (s_14)
    (s_12) edge node [pos=0.5, sloped, below]{$b^f$} (s_22)
    (s_22) edge node [pos=0.5, sloped, above]{$c$} (s_23)
    (s_1) edge node [pos=0.5, sloped, above]{$f_{b}$} (s_11)
    (s_2) edge node [pos=0.5, sloped, above]{$f_{b}$} (s_12)
    (s_3) edge node [pos=0.5, sloped, below]{$f_{b}$} (s_13)
    (s_4) edge node [pos=0.5, sloped, above]{$f_{b}$} (s_14);

	\end{tikzpicture}
	\caption{The DFA model of $G_i^{k,s}$.}
    \label{fig:giks}
	\vspace{-6.5mm}
\end{figure}
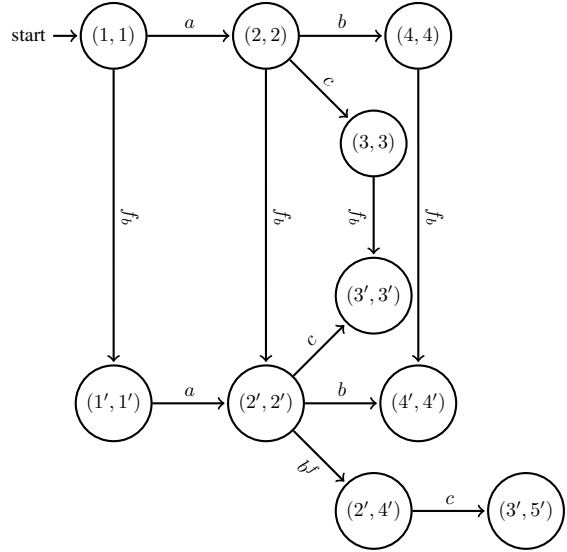

\end{example}
\begin{remark}
The construction procedure of $G_i^{k,s}$ is similar to that of the closed-loop system subject to ``sensor erasure attack" in \cite{carvalho2018detection}. Nevertheless, we use the sensor fault event to clearly distinguish the nominal part of the system from the faulty part. Furthermore, the procedure developed in \cite{carvalho2018detection} considers the worst-case scenario, whereas by introducing the sensor fault event, we can see from Example 2 that if a sensor fault takes place after the last occurrence of the corresponding sensor reading, the fault will pose no impact on the successive behaviors of the system.
\end{remark}

\subsection{SF-safe Controllability and Sensor Fault Tolerant Control}
By construction, $G_i^{k,s}$ $(i\in I)$ contains all the potential behaviors of the closed-loop subsystem subject to $f_{i,k}$ and $\sigma_{i,m}^f$. Two objectives need be fulfilled in order to achieve sensor fault tolerant control of $G_i^{k,s}$: (i) the occurrence of sensor fault event $f_{i,k}$ should be unambiguously determined before $G_i^{k,s}$ generates any unsafe behaviors; (ii) the fault-pruned controlled subsystem should be able to stop its evolution before generating any unsafe strings.

Let $\Psi(f_{i,k})=\{t\in L(G_i^{k,s})|t=t'f_{i,k},t'\in\Sigma_i^{k,s^*}\}$ denote the set of all strings in $L(G_i^{k,s})$ that end with the sensor fault event $f_{i,k}$. Let $P_{i,o}^k:\Sigma_i^{k,s^*}\to\Sigma_{i,o}^{k,s^*}$ be the post-fault observation projection. We introduce a variant of {\it safe controllability} \cite{paoli2011active,carvalho2018detection} in the context of active fault tolerant control, namely {\it SF-safe controllability}, to achieve the aforementioned control objectives for a fault-pruned subsystem.

\begin{definition}[SF-safe Controllability]\label{safecon}
Language $L(G_i^{k,s})$ is SF-safe controllable with respect to the projection $P_{i,o}^k$, the sensor fault event $f_{i,k}$ and the local safety property $L_i^{safe}$ if
\begin{equation}
\begin{split}
    &(\forall s_i\in\Psi(f_{i,k}))(\forall t_i\in L(G_i^{k,s})/s_i)\left[(s_it_i\not\models L_i^{safe})\land \right.\\
    &\left.(\forall s_i'\in\overline{s_it_i}-\{s_it_i\},\overline{s_i'}\models L_i^{safe})\right]\Rightarrow\mathcal{SC},
\end{split}
\end{equation}
where the safe controllability condition $\mathcal{SC}$ states as follows:
\begin{equation}\label{sfsafe}
    \begin{split}
    &\mathcal{SC}: (\exists t_{i,1},t_{i,2}\in\Sigma_i^{k,s^*},t_i=t_{i,1}t_{i,2})\\
    &\left[\left((\not\exists w_i\in L(G_i^{k,s}))[P_{i,o}^k(w_i)=P_{i,o}^k(s_it_{i,1})\land f_{i,k}^f\not\in w_i]\right)\land\right.\\
    &\left.(\Sigma_{i,c}^{k,s}\in t_{i,2})\right].
\end{split}
\end{equation}
\end{definition}

Intuitively, $L(G_i^{k,s})$ is SF-safe controllable if for any string $t_i$ in $L(G_i^{k,s})$ following the sensor fault event $f_{i,k}$ that may violate $L_i^{safe}$ (in the sense of Definition~\ref{satisfaction}), there exists: (i) a proper prefix $t_{i,1}$ of $t_i$ that assures the detection of $f_{i,k}$ before the fault-pruned subsystem generates any unsafe behavior (safe diagnosable); (ii) a locally controllable event after this prefix but still prior to the execution of the unsafe behavior (safe controllable). In other words, after the detection of the fault, unsafe behaviors can always be prohibited by disabling this locally controllable event. The above discussion is formally summarized as the following theorem, which asserts that SF-safe controllability is the necessary and sufficient condition for $G_i^{k,s}$ to ensure safety.

\begin{theorem}
The closed-loop subsystem $G_i^{k,s}$ subject to the sensor fault $f_{i,k}$ will not generate any safety-violating string (in the sense of Definition~\ref{satisfaction}) if and only if it is SF-safe controllable with respect to $P_{i,o}^k$, $f_{i,k}$ and $L_i^{safe}$.
\end{theorem}

The occurrence of the sensor fault event $f_{i,k}$ is determined by a {\it diagnoser}. The construction procedure of the diagnoser is presented in \cite{sampath1995diagnosability} and is omitted here. Before proceeding to the sensor fault tolerant control strategy for $G_i^{k,s}$, we first review the concept of first-entered certain states in the diagnoser \cite{paoli2011active}
\begin{definition}[First-entered Certain States]
Let $G_i^d=(Q_i^d,\Sigma_{i,o}^{k,s},\delta_i^d,q_{i,0}^d)$ be the diagnoser constructed for $G_i^{k,s}$ and $P_{i,o}^k$. Define $Q_i^{YN}=\{q\in Q_i^d\vert q \mbox{ is uncertain}\}$, $Q_i^N=\{q\in Q_i^d\vert q \mbox{ is normal}\}$ and $Q_i^Y=\{q\in Q_i^d\vert q \mbox{ is certain}\}$. The set of first-entered certain states is $\mathcal{FC}_i=\{q\in Q_i^Y\vert(\exists q'\in Q_i^N\cup Q_i^{YN},\sigma\in\Sigma_{i,o}^{k,s})[\delta_i^d(q',\sigma)=q]\}$.
\end{definition}

Let $Q_{i,B}=\{q\in Q_i^{k,s}\vert (\exists s\in L(G_i^{k,s}),\delta_i^{k,s}(q_{i,0}^{k,s},s)=q)[\overline{s}\not\models L_i^{safe}]\}$ denote the set of {\it unsafe} states in $G_i^{k,s}$. By introducing $Q_{i,B}$, $G_i^d$ can be modified as a {\it safe diagnoser} \cite{paoli2005safe} and the SF-safe controllability of $L(G_i^{k,s})$ can then be verified offline, as stated in the following proposition.
\begin{proposition}
Language $L(G_i^{k,s})$ is SF-safe controllable with respect to $P_{i,o}^k$, $f_{i,k}$ and $L_i^{safe}$ if and only if for the safe diagnoser $G_i^d$:
\begin{enumerate}[(i)]
    \item there does not exist a state $q_i^{YN}=\{(q_{i_1}^{k,s},l_{i_1}),(q_{i_2}^{k,s},l_{i_2}),$ $\ldots,(q_{i_K}^{k,s},l_{i_K})\}\in Q_i^{YN}$ such that $\exists j\in\{1,2,\ldots, K\}$, $l_{i_j}=Y$ but $q_{i_j}^{k,s}\in Q_{i,B}$;
    \item there does not exist a state $q_i^{Y}=\{(q_{i_1}^{k,s},Y),(q_{i_2}^{k,s},Y),$ $\ldots,(q_{i_K}^{k,s},Y)\}\in \mathcal{FC}_i$ such that $\exists j\in\{1,2,\ldots, K\}$, $q_{i_j}^{k,s}\in Q_{i,B}$;
    \item for any $q_i^{Y}=\{(q_{i_1}^{k,s},Y),(q_{i_2}^{k,s},Y),$ $\ldots,(q_{i_K}^{k,s},Y)\}\in \mathcal{FC}_i$ and $\forall j\in\{1,2,\ldots, K\}$, there does not exist a state $q_{i_{j'}}\in Q_{i,B}$ and a string $s\in \Sigma_{i,uc}^{k,s^*}$ such that $q_{i_{j'}}=\delta_i^{k,s}(q_{i_j},s)$.
\end{enumerate}
\end{proposition}
\begin{IEEEproof}
Conditions (i) and (ii) are the necessary and sufficient conditions for safe diagnosability and the proof is presented in \cite{paoli2005safe}. The necessity and sufficiency of Condition (iii) follow immediately from Definition~\ref{safecon}.
\end{IEEEproof}

\begin{example}
Let us revisit $G_i^{k,s}$ constructed in Example 2. From Fig.~\ref{fig:giks}, we can see that the observation $ac$ shall correspond to the diagnoser state $q_i=\{((3,3),N),((3',3'),Y),$ $((3',5'),Y)\}$, which is an $f_{i,k}$-uncertain state. On the other hand, since $(3',5')$ is an unsafe state, thus Condition (i) in Proposition 1 fails to be satisfied and thus $L(G_i^{k,s})$ is not SF-safe controllable.
\end{example}

\begin{figure}[h]
\begin{center}
    \centerline{\includegraphics[width=0.60\textwidth]{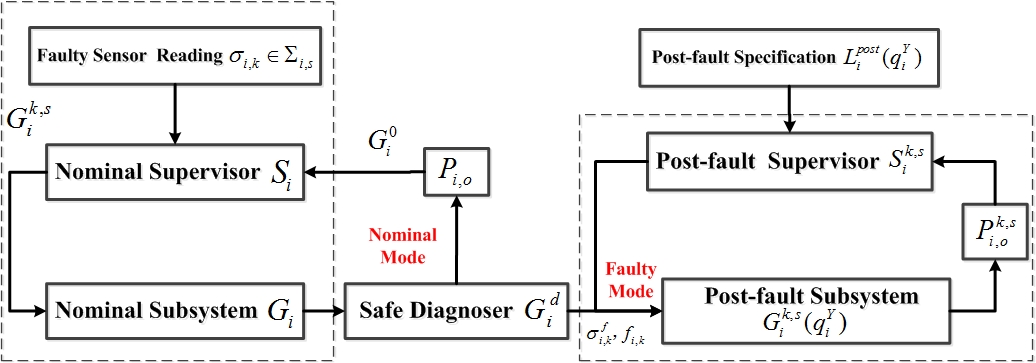}}
    \caption{Sensor fault tolerant control loop of $G_i^{k,s}$.}
    \label{fts}
  \end{center}
  \vspace{-7.5mm}
\end{figure}

We adopt an active approach to address the loss of the sensor reading $\sigma_{i,k}$, as shown in Fig.~\ref{fts}. Different from the actuator fault tolerant control, the sensor fault event $f_{i,k}$ is assumed to be locally unobservable and therefore, detection of $f_{i,k}$ should be performed by associating $G_i^{k,s}$ with the safe diagnoser $G_i^d$. If $L(G_i^{k,s})$ is safe controllable with respect to $P_{i,o}^k$, $f_{i,k}$ and $L_i^{safe}$, then any occurrence of the sensor fault $f_{i,k}$ can be determined by $G_i^d$ without generating any unsafe behaviors.

When $G_i^d$ reports that $G_i^{k,s}$ has evolved to a first-entered certain state $q_i^{Y}=\{(q_{i_1}^{k,s},$ $Y),(q_{i_2}^{k,s},Y),$ $\ldots,(q_{i_K}^{k,s},Y)\}\in \mathcal{FC}_i$, the post-fault {\it uncontrolled} subsystem $G_i^{k,s}(q_i^{Y})$ can be formed by disabling the nominal supervisor $S_i$ then taking the accessible part of of the uncontrolled faulty subsystem $G_i^k$ starting from each $q_{i_j}$, $j\in\{1,2,\ldots,K\}$. In order to make $G_i^{k,s}(q_i^Y)$ deterministic, we add a new initial state $q_{i,0}^Y$ and connect it with each $q_{i_j}$ with a transition labeled as $(q_{i,0}^Y,detect_j,q_{i_j})$, where $j\in\{1,2,\ldots,K\}$ and $detect_j$ is locally uncontrollable. The SF-safe controllability of $L(G_i^{k,s})$ indicates that there always exists a locally controllable event $\sigma_j$ corresponding to each $j$ that can be disabled to prevent the subsystem from generating an unsafe string after evolving to $q_{i_j}$. Let $t_{i_j,0}\in L(G_i^{k,s})$ be the string such that $\delta_i^{k,s}(q_{i,0}^{k,s},t_{i_j,0})=q_{i_j}$. We require that local safety can still be assured after the detection of the fault and the post-fault specification $L_i^{post}(q_i^Y)$ is given by
\begin{equation}\label{lipostqiy}
    L_i^{post}(q_i^Y)=\bigcup_{j=1}^K detect_j(L_i^{safe}/t_{i_j,0}),
\end{equation}
which turns to be a prefix-closed sublanguage of $L(G_i^{k,s}(q_i^Y))$. The following property of sensor fault tolerance is hence formally defined in terms of $G_i^{k,s}(q_i^{Y})$ and $L_i^{post}(q_i^Y)$.

\begin{definition}[Sensor Fault Tolerance]
Language $L(G_i^{k,s}(q_i^{Y}))$ is said to be {\it sensor fault tolerant} with respect to $f_{i,k}$ and $L_i^{post}(q_i^Y)$ if there exists a non-empty sublanguage of $L_i^{post}(q_i^Y)$ that is controllable with respect to $G_i^{k,s}(q_i^{Y})$ and $\Sigma_{i,uc}^{k,s}\cup\{detect_j\vert q_{i_j}\in q_i^Y\}$ and observable with respect to $G_i^{k,s}(q_i^{Y})$ and $\Sigma_{i,o}^{k,s}$. $L(G_i^{k,s})$ is said to be sensor fault tolerant if for all $q_i^Y\in\mathcal{FC}_i$, $L(G_i^{k,s}(q_i^{Y}))$ is sensor fault tolerant with respect to $L_i^{post}(q_i^Y)$.
\end{definition}



We present the following theorem that states the necessary and sufficient conditions for the sensor fault tolerance of $L(G_i^{k,s}(q_i^Y))$.
\begin{theorem}\label{stolerance}
Language $L(G_i^{k,s}(q_i^Y))$ is sensor fault tolerant with respect to $f_{i,k}$ and $L_i^{post}(q_i^Y)$ if and only if language $\inf C_i^{k,s}(\{\epsilon\})$, computed with respect to $L(G_i^{k,s}(q_i^Y))$ and $\Sigma_{i,uc}^{k,s}\cup\{detect_j\vert q_{i_j}\in q_i^Y\}$, satisfies that $\inf C_i^{k,s}(\{\epsilon\})\models L_i^{post}(q_i^Y)$.
\end{theorem}
\begin{IEEEproof}
The theorem can be proved in a similar way as that of Theorem~\ref{atolerance}. Note that we use $\models$ instead of $\subseteq$ in this theorem due to the introduction of the detection event $detect_j$ for each distinct $q_{i_j}\in q_i^Y$.
\end{IEEEproof}

Theorem~\ref{stolerance} in fact guarantees the existence of a satisfactory post-fault supervisor. Since the fault-pruned subsystem $G_i^{k,s}$ and the corresponding safe diagnoser $G_i^d$ can both be computed offline, all the possible post-fault transition diagram of $G_i$ after entering a state $q_i^Y\in \mathcal{FC}_i$ can also be obtained offline. Therefore, a post-fault supervisor $S_i^{k,s}(q_i^Y)$ starting from the states in $q_i^Y$ can be implemented online and the system shall switch to $S_i^{k,s}(q_i^Y)$ after $G_i^{k,s}$ visits any state in $q_i^Y$.

\begin{remark}
Although sensor fault tolerance of the subsystem $G_i$ can always be assured by synthesizing the nominal supervisor $S_i$ with respect to $\Sigma_{i,o}^{k,s}$ rather than $\Sigma_{i,o}$ in the first place; however, this approach would presumably lead the controlled subsystem to perform more restrictive behaviors.
\end{remark}

\subsection{Fault Tolerant Control with Multiple Faulty Sensors}
We now consider the case in which all sensor readings in $\Sigma_{i,s}$ may become faulty. For such a purpose, we first aim at constructing the DFA model of $G_i^0$, namely $G_i^{F,s}$, that characterizes the behaviors of $G_i^0$ in the presence of loss of sensor readings in $\Sigma_{i,s}$. Specifically, we first build the counterpart of $G_{i,k}$ in the multi-fault case as follows:
\begin{equation}
G_{i,F}=(Q_{i,F},\Sigma_{i,F},\delta_{i,F}),
\end{equation}
where $Q_{i,F}=\{q_{i,F,l}|q_{i,l}\in Q_i\}$ is a copy of $Q_i$, $\Sigma_{i,F}=\Sigma_i\cup\Sigma_{i,s}^F\cup\Sigma_{i,s}^f$, and the transition function $\delta_{i,F}$ is defined as follows: for any $q_{i,F,l}\in Q_{i,F}$ and $\sigma\in\Sigma_{i,F}$,
\begin{equation*}
    \delta_{i,F}(q_{i,F,l},\sigma)=
    \begin{cases}
    q_{i,F,l'}, \mbox{ if } &\sigma\not\in\Sigma_{i,s}^f\land\delta_i(q_{i,l},\sigma)=q_{i,l'};\\
    q_{i,F,l'}, \mbox{ if } &\sigma=\sigma_{i,k}^f\in\Sigma_{i,s}^f\land\\
    &\delta_i(q_{i,l},\sigma_{i,k})=q_{i,l'}.
    \end{cases}
\end{equation*}
Therefore, similar to \eqref{gikmodel}, the DFA model of $G_i$ in the presence of multiple sensor faults can be obtained as
\begin{equation}\label{gifmodel}
G_i^F=(Q_i^F,\Sigma_i^F,\delta_i^F,q_{i,0}^F),
\end{equation}
where $Q_i^F=Q_i\cup Q_{i,F}$, $\Sigma_i^F=\Sigma_{i,F}$, $q_{i,0}^F=q_{i,0}$, with the transition function $\delta_i^F=\delta_i\cup\delta_{i,F}\cup\{(q_{i,l},f_{i,k},q_{i,k,l})\vert q_{i,l}\in Q_i,f_{i,k}\in\Sigma_{i,s}^F\}$.

On the other hand, the DFA model $S_i^F$ of $S_i$ in the multi-fault is computed, and we have
\begin{equation*}
S_{i,F}=(X_{i,F},\Sigma_{i,F},\xi_{i,F}),
\end{equation*}
where $X_{i,F}=\{x_{i,F,l}\vert x_{i,l}\in X_i\}$, and for any $x_{i,F,l}\in X_{i,F}$ and $\sigma\in \Sigma_{i,F}$, $\xi_{i,F}(x_{i,F,l},\sigma)$ is formally defined as
\begin{equation*}
    \xi_{i,F}(x_{i,F,l},\sigma)=
    \begin{cases}
    x_{i,F,l'}, &\mbox{if } \xi_i(x_{i,l},\sigma)=x_{i,l'}; \\
    x_{i,F,l}, &\mbox{if } [\sigma=\sigma_{i,k}^f\in\Sigma_{i,s}^f\land\\
    &\xi_i(x_{i,l},\sigma_{i,k})!]\lor[\sigma\in\Sigma_{i,uc}\cup\Sigma_{i,s}^f  \\
               & \land\neg\xi_i(x_{i,l},\sigma)!].
    \end{cases}
\end{equation*}
The unified model of $S_i^F$ is hence obtained as
\begin{equation}\label{sifmodel}
S_i^F=(X_i^F,\Sigma_i^F,\xi_i^F,x_{i,0}^F),
\end{equation}
where $X_i^F=X_i\cup X_{i,F}$, $\Sigma_i^F=\Sigma_{i,F}$, $x_{i,0}^F=x_{i,0}$ and $\xi_i^F=\xi_i\cup\xi_{i,F}\cup\{(x_{i,l},f_{i,k},x_{i,F,l})\vert x_{i,l}\in X_i,f_{i,k}\in\Sigma_{i,s}^F\}$. Finally, the closed-loop model $G_i^{F,s}$ is computed in a similar way as \eqref{giks}
\begin{equation}\label{gifs}
    G_i^{F,s}=S_i^F\| G_i^F:=(Q_i^{F,s},\Sigma_i^{F,s},\delta_i^{F,s},q_{i,0}^{F,s}),
\end{equation}
where the set of post-fault local events is given by $\Sigma_i^{F,s}=\Sigma_{i,F}$ and can be further partitioned as
\begin{equation}\label{sensoreventmulti}
\begin{split}
&\Sigma_{i,c}^{F,s}=\Sigma_{i,c},\Sigma_{i,uc}^{F,s}=\Sigma_{i,uc}\cup\Sigma_{i,s}^f\cup\Sigma_{i,s}^F;\\
&\Sigma_{i,o}^{F,s}=\Sigma_{i,o},\Sigma_{i,uo}^{F,s}=\Sigma_{i,uo}\cup\Sigma_{i,s}^f\cup\Sigma_{i,s}^F.
\end{split}
\end{equation}

\begin{figure}[h]
\begin{center}
    \centerline{\includegraphics[scale=0.45]{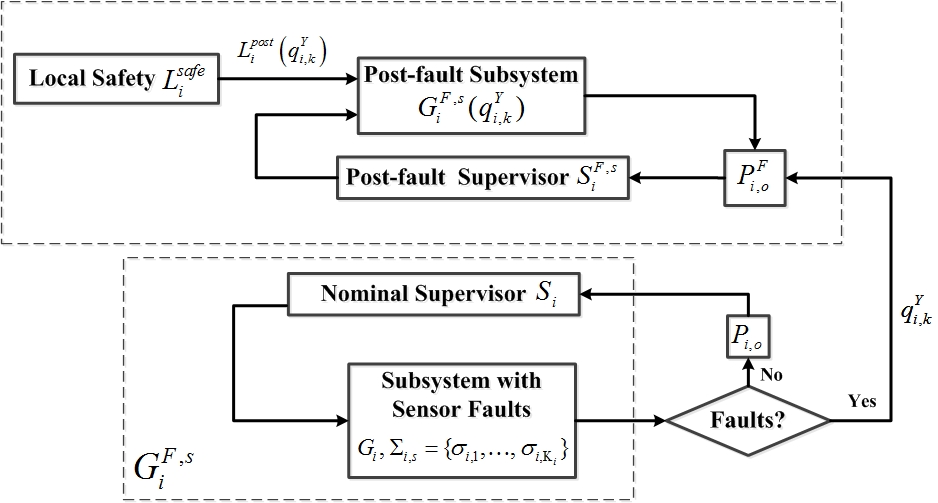}}
    \caption{Sensor fault tolerant control with multiple faults.}
    \label{ftsm}
  \end{center}
  \vspace{-9.0mm}
\end{figure}

A sensor fault tolerant control framework is developed in Fig.~\ref{ftsm} to resolve the impacts of multiple sensor faults, where $P_{i,o}^F:\Sigma_i^{F,s^*}\to\Sigma_{i,o}^{F,s^*}$ stands for the post-fault observation projection in the multi-fault case. In the multi-fault case, the safe diagnoser $G_i^d$ can be modified to distinguish different sensor faults by introducing fault labels corresponding to each sensor fault event. When no sensor fault is detected by $G_i^d$, the subsystem $G_i^{F,s}$ remains in the nominal mode. If for each $k\in\{1,2,\ldots,K_i\}$, the language $L(G_i^{F,s})$ is SF-safe controllable with respect to $P_{i,o}^F$, $f_{i,k}$ and $L_i^{safe}$, the safe diagnoser $G_i^d$ is able to correctly detect the occurrence of $f_{i,k}$ before $G_i^{F,s}$ generates any (locally) unsafe behaviors. With slightly abusing the notations, we still denote by $\mathcal{FC}_i$ the set of first-entered certain states with respect to all possible sensor fault events in $\Sigma_{i,s}^F$. Thanks to $\mathcal{FC}_i$, we can apply Proposition~1 for the verification of the SF-safe controllability of $L(G_i^{F,s})$ in the presence of multiple sensor faults.

When entering a certain state $q_i^Y\in\mathcal{FC}_i$, the diagnoser reports the occurrence of the corresponding sensor fault $f_{i,k}$ for some $k\in\{1,2,\ldots,K_i\}$ and interrupts the operation of the nominal supervisor $S_i$. In response to the detection of the sensor fault, the construction procedure for the post-fault uncontrolled subsystem in the single-fault case can be inherited to compute $G_i^{k,s}(q_i^Y)$, while the post-fault control specification is obtained as \eqref{lipostqiy}. The solvability of the fault tolerant control problem under multiple sensor faults is presented as follows.
%
%

\begin{theorem}\label{multisen}
There exists a post-fault supervisor $S_i^{F,s}(q_i^Y)$ after $G_i^d$ evolves to a certain state $q_i^Y$ that drives the post-faul subsystem $G_i^{F,s}(q_i^Y)$ to satisfy $L_i^{post}(q_i^Y)$ \eqref{lipostqiy} under arbitrary order of loss of sensor readings in $\Sigma_{i,s}$ if and only if language $\inf C_i^{F,s}(\{\epsilon\})$, computed with respect to $L(G_i^{F,s}(q_i^Y))$ and $\Sigma_{i,uc}^{F,s}\cup\{detect_j\vert q_{i_j}\in q_i^Y\}$, satisfies that $\inf C_i^{F,s}(\{\epsilon\})\models L_i^{post}(q_i^Y)$.
\end{theorem}

\begin{IEEEproof}
The proof is similar to that of Theorem~\ref{atolerance} and is omitted here.
\end{IEEEproof}

The notion $S_i^F\|G_i^F$ is also adopted here to represent the overall closed-loop model of the subsystem $G_i$ $(i\in I)$ subject to multiple sensor faults. The following theorem states that the active fault tolerant architecture proposed in Fig.~\ref{ftsm} shall ensure the local safety of $G_i$ regardless of faults.
\begin{theorem}\label{thm:fts}
Consider the subsystem $G_i$ $(i\in I)$ subject to possible loss of sensor readings $\sigma_{i,1},\sigma_{i,2},\ldots,\sigma_{i,K_i}$. If for each $k\in\{1,2,\ldots,K_i\}$, $L(G_i^{k,s})$ is SF-safe controllable with respect to $P_{i,o}^F$, $f_{i,k}$ and $L_i^{safe}$ and for each $q_i^Y\in\mathcal{FC}_i$, $L(G_i^{k,s}(q_i^Y))$ is sensor fault tolerant with respect to $f_{i,k}$ and $L_i^{post}(q_i^Y)$, then the nominal supervisor $S_i$ and the post-fault supervisor $S_i^{F,s}$ will jointly enforce the fulfillment of local safety requirement $L_i^{safe}$, i.e., $S_i^F\|G_i^F\models L_i^{safe}$.
\end{theorem}

\begin{IEEEproof}
By definition, the behaviors $L(S_i^F\|G_i^F)$ of the subsystem $G_i$ in the presence of multiple sensor faults should be the concatenation of three parts: the nominal behaviors of $G_i$ under the supervision of $S_i$ before the occurrence of the fault, the behaviors generated after the occurrence but before the detection of the fault, and the behaviors generated by the post-fault subsystem controlled by $S_i^{F,s}$. More specifically, $L(S_i\|G_i)$ should be represented as the following form:
\begin{equation*}
L\left(S_i^F\|G_i^F\right)=\left\{\overline{s_if_{i,k}t_idetect_jt'_i}\right\}, \\
\end{equation*}
where $s_i\in L(G_i^0)$ is the string generated before $f_{i,k}$, $t_i\in L(G_i^{k,s})\cap\Psi(f_{i,k})$ is the string executed by $G_i^{k,s}$ in the faulty mode but before $f_{i,k}$ is detected, and $t'_i$ is the string generated by the post-fault controlled subsystem $S_i^{F,s}(q_i^Y)\|G_i^{F,s}(q_i^Y)$ if $G_i^d$ enters a certain state $q_i^Y$. In other words, $\overline{s_i}\subseteq L_i\subseteq L_i^{safe}$ and $s_if_{i,k}t_i\models L_i^{safe}$ can be guaranteed due to the SF-safe controllability of $L(G_i^{k,s})$ for each $k\in\{1,2,\ldots,K_i\}$. Furthermore, when evolving to any fault certain state $q_i^Y$, the sensor fault tolerance property can assure the existence of a post-fault supervisor $S_i^{F,s}(q_i^Y)$ that satisfies $L_i^{post}(q_i^Y)$ in the faulty mode in the presence of $\Sigma_{i,uc}^{F,s}$ and $\Sigma_{i,uo}^{F,s}$; therefore, recall \eqref{lipostqiy}, we can write that
\begin{equation*}
    \overline{detect_jt'_i}\subseteq L_i^{post}(q_i^Y)/(s_it_i)
\end{equation*}
for some $j$ such that there exists $q_{i_j}\in q_i^Y$. Let $P_i^{F,s}$ denote the natural projection from $[\Sigma_i^F\cup\{detect_j\vert q_{i_j}\in q_i^Y\}]^*$ to $\Sigma_i^*$.
 It then follows that
\begin{equation}
P_i^{F,s}\left[L(S_i^F\|G_i^F)\right]=\left\{\overline{s_it_it'_i}\right\}\subseteq L_i^{safe},
\end{equation}
which is equivalent to $S_i^F\|G_i^F\models L_i^{safe}$.
\end{IEEEproof}

\subsection{Supervisory Control with Combinations of Faults}
So far, we have only considered one singe type of faults that may occur in a subsystem. In this subsection, we aim at extending the proposed fault tolerant control approaches to take both actuator and sensor faults into consideration. Without loss of generality, we consider a subsystem $G_i$ $(i\in I)$ in which both $\Sigma_{i,s}=\{\sigma_i\}$ and $\Sigma_{i,a}=\{\eta_i\}$ are singletons. We assume that the sensor fault occurs before the actuator fault, and the fault tolerance of the supervisory control strategy of $G_i$ is sketched as follows. Note that our methodology can be generalized to other combinations of faults.
(1) When no fault is detected in $G_i$, the nominal supervisor $S_i$ is employed such that $L(G_i^0)=L_i\subseteq L_i^{safe}$. With $\Sigma_{i,s}^F=\{f_i\}$ and $\Sigma_{i,s}^f=\{\sigma_i^f\}$, we can construct the fault-pruned model $G_i^{F,s}$ of $G_i^0$. Furthermore, $G_i^{F,s}$ is monitored by the safe diagnoser $G_i^d$.

(2) If $L(G_i^{F,s})$ is SF-safe controllable and $G_i^d$ detects the occurrence of the sensor fault $f_i$ by entering a certain state $q_i^Y\in\mathcal{FC}_i$, the operation of the nominal supervisor $S_i$ can then be disabled before generating any unsafe behaviors. Facing the post-fault model $G_i^{F,s}(q_i^Y)$ of the uncontrolled subystem and the post-fault specification $L_i^{post}(q_i^Y)$, the sensor fault tolerant supervisor $S_i^{F,s}(q_i^Y)$ can be synthesized and implemented online, provided that $L(G_i^{F,s}(q_i^Y))$ is sensor fault tolerant with respect to $f_i$ and $L_i^{post}(q_i^Y)$. After switching to $S_i^{F,s}$, we denote by $S_i^F\|G_i^F$ the overall closed-loop subsystem.

(3) We further assume that $t_i\in L(S_i^F\|G_i^F)$ is generated when the actuator fault event $h_i$ is detected. In this case, we set $G_i^{F^{suf}}(t_i)$ as the uncontrolled post-fault plant and the post-fault specification is updated as $L_i^{post}=L_i^{safe}/P_i^{F,s}(t_i)$ as the post-fault specification, where $P_i^{F,s}$ is defined in Theorem~\ref{thm:fts}. In this faulty mode, we update the local event sets as \eqref{aevent1} and \eqref{aevent2} accordingly, and if actuator fault tolerance of $L(G_i^{F^{suf}}(t_i))$ is satisfied, we can switch to a second post-fault supervisor $S_i^{F,a}$ that assures local safety with the faulty actuator and sensor.

For brevity of presentation, we still use $S_i^F\|G_i^F$ $(i\in I)$ as a unified notation to represent the closed-loop subsystem in the presence of combinations of faults. The following theorem, suggesting that the integration of the fault tolerant control techniques jointly enforce the local safety of a subsystem subject to faults, can be viewed as an immediate result by following the conclusions of Theorem~\ref{thm:fta} and Theorem~\ref{thm:fts}.

\begin{theorem}\label{thm:ft}
For the subsystem $G_i$ $(i\in I)$ that is subject to actuator faults in $\Sigma_{i,a}^F$ and sensor faults in $\Sigma_{i,s}^F$, the proposed fault tolerant techniques depicted in Fig.~\ref{fta} and Fig.~\ref{ftsm} will result in $S_i^F\|G_i^F\models L_i^{safe}$.
\end{theorem}

\section{Assume-guarantee Post-fault Coordination of Distributed DESs}\label{sec:discussion}
Theorem~\ref{thm:ft} guarantees that local safety of each subsystem can be enforced after switching to post-fault supervisor(s). Nevertheless, undesirable behaviors may still arise when post-fault subsystems are coordinated with the nominal ones, leading to the violation of the global specification $L$. We resolve this concern in this section by developing fault tolerant coordination strategies among the subsystems . In particular, an assume-guarantee paradigm
\cite{cobleigh2003learning} is exploited to efficiently refine the local supervisors in order to achieve the global specification.
\subsection{Essentials of Compositional Verification}
We first review the completion DFAs in the compositional verification procedures that are presented in \cite{cobleigh2003learning}.

\begin{definition}[Completion]\label{complement DFA}
Given $G=(Q,\Sigma,\delta,q_0,Q_m)$ with an ``error" state $q_e\not\in Q$, the completion of $G$ is defined as a DFA $\tilde{G}=(\tilde {Q},\Sigma,\tilde{\delta},q_0,Q_m)$ with $\tilde{Q}=Q\cup \{q_e\}$, and
$$
\forall \tilde{q}\in \tilde {Q}, \sigma\in \Sigma, \tilde {\delta}(\tilde{q},\sigma)=
\begin{cases}
\delta(\tilde{q},\sigma), & \mbox{if }\tilde {q}\in Q \land \delta(\tilde {q},\sigma)!, \\
q_e, & \mbox{otherwise. }
\end{cases}
$$
\end{definition}

It can be shown that $L(\tilde{G})=\Sigma^*$ and $L_m(\tilde{G})=L_m(G)$. The {\it complement} of a DFA $G$ over $\Sigma$, written as $coG$, is a DFA such that $L_m(coG):=\Sigma^*-L_m(G)$ and can be constructed by swapping the marked states of $\tilde G$ with its non-marked states and vice versa. Recall Definition~\ref{satisfaction}, it is shown in \cite{cobleigh2003learning} that a system $M$ violates a property $P$ if and only if the error state $q_e$ is reachable in $M\|\tilde P$, where $\tilde P$ is the completion of $P$.

An assume-guarantee formula is a triple $\langle A \rangle M \langle P \rangle $, where $M$ is the system, $P$ is the property to be verified and $A$ is an assumption about $M$'s environment, each of which is represented by a corresponding DFA. The formula holds if whenever $M$ is part of a system satisfying $A$, the system must also guarantee the property $P$, i.e., $\forall E$, $E\| M \models A$ implies that $E\|M\models P$ \cite{puasuareanu2008learning}. It is shown in \cite{puasuareanu2008learning} that when $\Sigma_P\subseteq \Sigma_A\cup\Sigma_M$, $\langle A \rangle M \langle P \rangle$ if and only if $q_e$ is unreachable in $A\|M\|\tilde P$, i.e., $M\|A\models P$. A series of symmetric and asymmetric proof rules are incorporated for the assume-guarantee paradigm. When $M=\|_{i\in I} M_i$ is a system that is composed of $n$ components, the following symmetric proof rule SYM-N \cite{puasuareanu2008learning} is adopted:
\begin{center}
\begin{tabular}{ll}
$i=1,\ldots,n$ & $\langle A_i \rangle M_i \langle P \rangle $ \\
$i=n+1$ & $L_m(coA_1\| coA_2 \| \cdots \| coA_n)\subseteq L(P)$ \\
\hline
 & $\langle {\rm true} \rangle M_1\| M_2 \| \cdots \| M_n \langle P \rangle$
\end{tabular}
\end{center}
where $A_i$ is the assumption about $M_i$'s environment and $coA_i$ is the complement of $A_i$. For $i\in I$, we require that $\Sigma_P\subseteq \bigcup_{i\in I}\Sigma_i=\Sigma$ and $\Sigma_{A_i}\subseteq \left(\bigcap_{i\in I} \Sigma_i\right)\cup\Sigma_P$. It is shown that the SYM-N proof rule is sound and complete \cite{puasuareanu2008learning}.

The assumption $A$ in an assume-guarantee formula need not be unique, and we are particularly interested in the {\it weakest} assumption about a system's environment. Formally, when $n=2$, the weakest assumption is defined as follows.
\begin{definition}[Weakest Assumption]\label{weakest assumption}
\cite{puasuareanu2008learning} Let $M_1$ and $M_2$ be two system components defined over $\Sigma_1$ and $\Sigma_2$, respectively, and $P$ be a property defined over $\Sigma_P$. Let $\Sigma_A:=(\Sigma_1\cup\Sigma_P)\cap\Sigma_2$ be an interface alphabet, the weakest assumption for $M_1$ is a DFA $A_w$ defined over $\Sigma_A$ such that for any component $M_2$, $\langle {\rm true} \rangle M_1\|P_A(M_2) \langle P \rangle $ if and only if $\langle {\rm true} \rangle M_2 \langle A_w \rangle $, where $P_A$ denotes the natural projection from $\Sigma_2^*$ to $\Sigma_A^*$.
\end{definition}

For each $i\in I$, let $\Sigma_{-i}:=\bigcup_{j\in I-\{i\}} \Sigma_j$ denote the set of events that belong to all the subsystems except $G_i$. By setting $M_{-i}=\|_{j\in I-\{i\}} M_j$ and $\Sigma_{A_i}=(\Sigma_i\cup\Sigma_P)\cap\Sigma_{-i}$, the weakest assumption $A_i$ with respect to $M_i$, $M_{-i}$ and $P$ can be constructed according to Definition~\ref{weakest assumption}. Note that the number of states of $A_i$ is generally less than the number of states of $M_i$; therefore, deployment of the assume-guarantee reasoning can efficiently justify $\|_{i\in I} M_i\models P$ by avoiding the computation of the parallel composition $\|_{i\in I} M_i$.

\subsection{Fault Tolerant Coordination of Distributed DESs}
In this subsection, we propose an assume-guarantee scheme to coordinate nominal subsystems with subsystems after switching to post-fault supervisor(s). To apply the SYM-N proof rule for the coordination of the controlled distributed DES, we use $L\subseteq \Sigma^*$ as the property to be verified and let $M_i:=G_i^0$ $(i\in I)$ denote component module of the controlled subsystem in the nominal mode. In this case, the weakest assumption $A_i$ with respect to $M_i$ and $L$ is a DFA that is defined over $\Sigma_{A_i}:=\Sigma_{-i}$. Since the SYM-N rule is sound and complete, all the assumptions $A_i$ $(i\in I)$ jointly satisfy the $(n+1)$-th premise of the SYM-N proof rule.

For the sake of simplicity of presentation, we assume that operation of one subsystem $M_i$ $(i\in I)$ suffers from possible actuator/sensor faults and switches to a post-fault supervisor. The post-fault counterpart of $M_i$, written as $M_i^F$, is defined as a DFA that satisfies $L(M_i^F)=P_i^F\left[L(S_i^F\|G_i^F)\right]\subseteq \Sigma_i^*$, where $P_i^F$ denotes the natural projection from $\Sigma_i^{F^*}$ to $\Sigma_i^*$. According to Definition~\ref{satisfaction}, $\left[\|_{j\in I-\{i\}}(S_j\|G_j)\right]\|(S_i^F\|G_i^F) \models L$ reduces to $\left(\|_{j\in I-\{i\}} M_j\right)\|M_i^F \models L$. The weakest assumption with respect to the post-fault subsystem module $M_i^F$ and the global specification $L$, written as $A_i^F$, is also defined over $\Sigma_{-i}$ and can be computed accordingly via existing methods \cite{cobleigh2003learning}.

\begin{figure}[h]
\begin{center}
    \centerline{\includegraphics[width=0.60\textwidth]{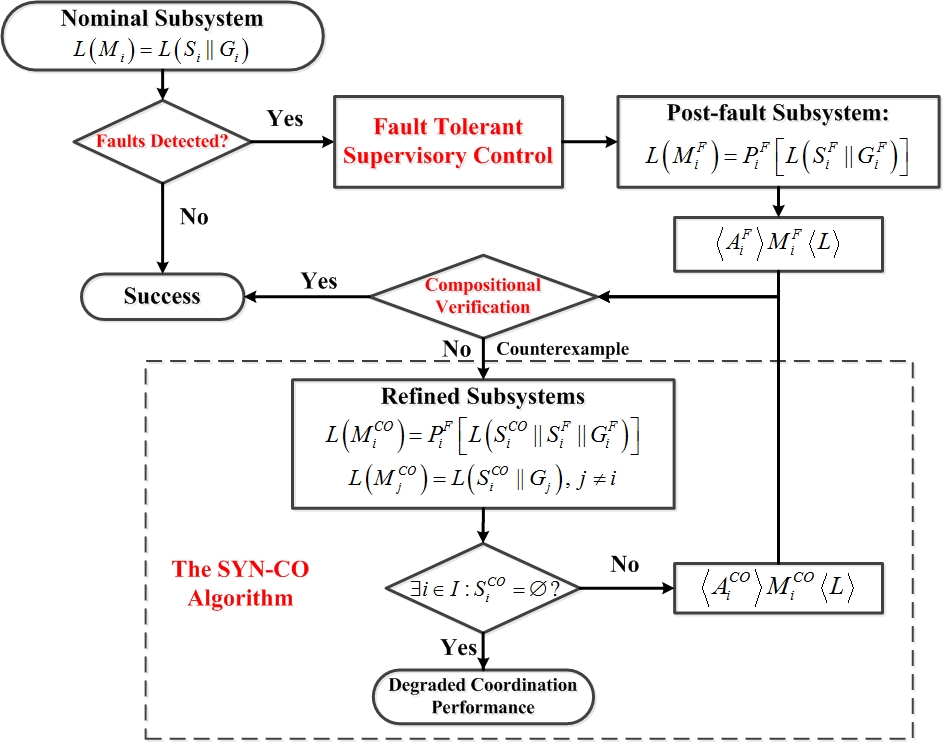}}
    \caption{Fault tolerant assume-guarantee coordination.}
    \label{agfmulti}
  \end{center}
 \vspace{-9.5mm}
\end{figure}

The assume-guarantee paradigm for the post-fault coordination among the subsystems of $G$ is illustrated in Fig.~\ref{agfmulti}. The proposed coordination scheme first seeks to maintain the global specification even in the presence of a subsystem with post-fault local supervisor(s). The following proposition states the necessary and sufficient condition under which the successful coordination can be ensured without further supervisor refinement.

\begin{proposition}\label{limif}
$\left(\|_{j\in I-\{i\}} M_j\right)\|M_i^F \models L$ if and only if $L_m\left[(\|_{j\in I-\{i\}} coA_j)\|coA_i^F\right]\subseteq L$.
\end{proposition}
\begin{IEEEproof}
On the one hand, defined over $\Sigma_{-i}$, $A_i^F$ is the weakest assumption such that the $i$-th premise of the SYM-N proof rule, $\langle A_i^F \rangle M_i^F \langle L \rangle$, is satisfied. Furthermore, for each $j\in I-\{i\}$, the $j$-th premise of the SYM-N rule, $\langle A_j \rangle M_j \langle L \rangle$, holds automatically. If $L_m\left[(\|_{j\in I-\{i\}} coA_j)\|coA_i^F\right]\subseteq L$, the $(n+1)$-th premise of the SYM-N rule is satisfied and $\left(\|_{j\in I-\{i\}} M_j\right)\|M_i^F \models L$ is enforced by the soundness of the SYM-N proof rule.

On the other hand, if $\left(\|_{j\in I-\{i\}} M_j\right)\|M_i^F \models L$ holds, then for the weakest assumptions $A_j$ (with respect to $M_j$ $(j\in I-\{i\})$ and $L$) and $A_i^F$ (with respect to $M_i^F$ and $L$), premises $1$ through $n$ of the SYM-N proof rule are satisfied. Therefore, serving as the $(n+1)$-th premise of the SYM-N proof rule, $L_m\left[(\|_{j\in I-\{i\}} coA_j)\|coA_i^F\right]\subseteq L$ is assured by the completeness of the SYM-N proof rule.
\end{IEEEproof}

If Proposition~\ref{limif} cannot be satisfied, we need to figure out how to refine the controlled behaviors of both the nominal and faulty subsystem equipped with the post-fault supervisor. The coordination architecture shown in Fig.~\ref{agfmulti} implements the refinement by associating each subsystem with a {\it coordination} supervisor $S_i^{CO}$ $(i\in I)$. For the nominal subsystems, we reconfigure the control policies of the nominal subsystems and define $M_j^{CO}=S_j^{CO}||G_j$ $(j\ne i)$ as the component module; whereas the component module $M_i^{CO}$ of the post-fault subsystem $S_i^F\|G_i^F$ is defined as a DFA over $\Sigma_i^F$ such that $L(M_i^{CO})=P_i^F\left[L(S_i^{CO}\|S_i^F\|G_i^F)\right]$. Let $A_i^{CO}$ and $A_j^{CO}$ denote the weakest assumptions (with respect to $L$) for $M_i^{CO}$ and $M_j^{CO}$ $(j\ne i)$, respectively. Furthermore, we assume an ``infimally permissive" supervisor $S_i^{\inf}$ $(i\in I)$ to realize either $\inf C_j(\{\epsilon\})$ (computed with respect to $G_j$ and $\Sigma_{j,uc}$ $(j\ne i)$) or $\inf C_i^F(\{\epsilon\})$ (computed with respect to $S_i^F\|G_i^F$ and $\Sigma_{i,uc}^F$). We refer to $M_i^{\inf}$ and $A_i^{\inf}$ $(i\in I)$ as the component module and the weakest assumption DFAs when the coordination supervisor $S_i^{CO}$ reduces to $S_i^{\inf}$, respectively. The following theorem derives the necessary and sufficient condition for the existence of the $S_i^{CO}$'s in terms of $A_i^{\inf}$ $(i\in I)$.

\begin{theorem}\label{limifree}
There exists a coordination supervisor $S_i^{CO}$ for each subsystem $G_i$ $(i\in I)$ such that $\|_{i\in I} M_i^{CO}\models L$ if and only if $L_m\left(\|_{i\in I} coA_i^{\inf}\right)\subseteq L$.
\end{theorem}

\begin{IEEEproof}
Suppose that $L_m\left(\|_{i\in I} coA_i^{\inf}\right)\subseteq L$, where $A_i^{\inf}$ is the weakest assumption with respect to $M_i^{\inf}$ and $L$. The soundness of the SYM-N proof rule implies that $\|_{i\in I} M_i^{CO} \models L$, which suggests that $S_i^{\inf}$ $(i\in I)$ suffice to be a satisfactory coordination superviso.

Conversely, suppose that there exists a coordination supervisor $S_i^{CO}$ for either $S_i^F\|G_i^F$ or $G_j$ $(j\ne i)$ to jointly satisfy the global specification $L$. By definition, $L(M_i^{\inf})\subseteq L(M_i^{CO})$ always holds for each $i\in I$. Hence, we can write that
\begin{equation*}
\begin{split}
L(M_i^{\inf}\|A_i^{CO})&=L(M_i^{\inf})\|L(A_i^{CO})\\
&\subseteq L(M_i^{CO})\|L(A_i^{CO})\subseteq L,
\end{split}
\end{equation*}
which implies that $\langle A_i^{CO} \rangle M_i^{\inf} \langle L \rangle$; that is, $A_i^{CO}$ is an appropriate assumption with respect to $M_i^{\inf}$ and $L$. Since $A_i^{\inf}$ is the weakest assumption, $A_i^{CO}$ is stronger than $A_i^{\inf}$ and thus $L(A_i^{CO})\subseteq L(A_i^{\inf})$. Furthermore, since all the states in $A_i^{CO}$ and $A_i^{\inf}$ are marked, $L(A_i^{CO})\subseteq L(A_i^{\inf})$ implies that $L_m(coA_i^{\inf})\subseteq L_m(coA_i^{CO})$. Therefore, we have that
\begin{equation*}
\begin{split}
&L_m\left(\|_{i\in I} coA_i^{\inf}\right)= \|_{i\in I} L_m(coA_i^{\inf})\\
&\subseteq\|_{i\in I} L_m(coA_i^{CO}) = L_m\left(\|_{i\in I} coA_i^{CO}\right)\subseteq L
\end{split}
\end{equation*}
where the last inclusion is enforced by the satisfaction of $\|_{i\in I} M_i^{CO}\models L$ and the completeness of the SYM-N rule. The proof is hence completed.
\end{IEEEproof}

The intuition behind Theorem~\ref{limifree} states that if the infimally feasible behaviors performed by each subsystem in the fault operation cannot jointly maintain the global specification, there is no other way to achieve a successful coordination.

Starting from the component modules $M_j$ $(j\ne i)$ and $M_i^F$, the coordination procedure shown in Fig.~\ref{agfmulti} states as follows:

(1) Following the detection of actuator/sensor faults, the subsystem $G_i$ switches to the post-fault supervisor $S_i^F$ to satisfy the local safety requirement $L_i^{safe}$ and the post-fault component module $M_i^F$ is computed. With $M_i^F$ and $L$, the weakest assumption $A_i^F$ can be obtained and the satisfaction of Proposition~\ref{limif} is justified. If Proposition~\ref{limif} is satisfied, the fulfillment of $L$ is still maintained and no supervisor refinement is required.

(2) If Proposition~\ref{limif} fails to be satisfied, we apply Theorem~\ref{limifree} to determine whether or not the global specification can be accomplished by synthesizing appropriate coordination supervisor for each subsystem. If the coordination supervisors do exist, we compute the coordination supervisor $S_i^{CO}$ $(i\in I)$ for $S_i^F\|G_i^F$ and $G_j$ $(j\ne i)$ by applying the following counterexample-guided synthesis algorithm named SYN-CO\footnote{SYN-CO stands for ``synthesis for coordination}.

\begin{algorithm}[h]
\caption{The SYN-CO Algorithm}
\begin{algorithmic}[1]
\REQUIRE $M_i^F$, $G_j$ $(j\ne i)$, $\Sigma_i^F$, $\Sigma_j$, and $L$
\ENSURE $M_i^{CO}$ and $S_i^{CO}$ $(i\in I)$
\STATE Initialization: $M_i^{CO} \gets M_i^F$, $M_j^{CO} \gets M_j$ $(j\ne i)$
\STATE Compute $A_i^{CO}$ such that $\langle A_i^{CO} \rangle M_i^{CO}
\langle L \rangle$ $(i\in I)$
\IF {$L_m\left(\|_{i\in I} coA_i^{CO}\right)\subseteq L$}
\RETURN $S_i^{CO}$ and $M_i^{CO}$ $(i \in I)$
\ELSE
\STATE $M_i^{CO} \gets M_i^F$, $M_j^{CO} \gets G_j$ $(j\ne i)$
\STATE Compute $A_i^{CO}$ such that $\langle A_i^{CO} \rangle M_i^{CO}
\langle L \rangle$ $(i\in I)$
\ENDIF
\WHILE {$L_m\left(\|_{i\in I} coA_i^{CO}\right)\not\subseteq L$}
\STATE A counterexample $c\in\Sigma^*$ is returned by the compositional verification procedure
\STATE $L^{temp}_i \gets L(M_i^{CO})-P_i(c)$ $(i\in I)$
\STATE $L_i^{CO} \gets L^{temp}_i - (\Sigma_i^*-L^{temp}_i)\Sigma_i^*$ $(i\in I)$
\STATE Compute a maximally permissive coordination supervisor $S_i^{CO}$ such that $S_i^{CO}\|S_i^F\|G_i^F \models L_i^{CO}$
\STATE Compute a maximally permissive coordination supervisor $S_j^{CO}$ such that $S_j^{CO}\|G_j \models L_j^{CO}$ $(j\ne i)$
\STATE Update $M_i^{CO}$ $(i \in I)$
\STATE Compute $A_i^{CO}$ with respect to $M_i^{CO}$ and $L$
\ENDWHILE
\RETURN $S_i^{CO}$ and $M_i^{CO}$ $(i \in I)$
\end{algorithmic}
\end{algorithm}

The working procedure of the SYN-CO algorithm is explained as follows. First, $M_i^{CO}$ and $M_j^{CO}$ $(j\ne i)$ are initialized to be $M_i^F$ and $M_j$ (lines 1 and 2), respectively. Whenever $L_m\left(\|_{i\in I} coA_i^{CO}\right)\subseteq L$ holds, the fulfillment of $L$ is assured automatically by Proposition~\ref{limif} (lines 3 and 4). Otherwise, we aim at the synthesis of coordination supervisors and set $M_j^{CO}$ to be $G_j$ for the nominal subsystems (lines 5 to 8). If $L_m\left(\|_{i\in I} coA_i^{CO}\right)\not\subseteq L$, then a counterexample $c\in\Sigma^*$ that causes $\|_{i\in I} M_i^{CO}$ to violate $L$ is returned by the assume-guarantee compositional verification procedure (lines 9 and 10). The counterexample is utilized to generate new local specification $L_i^{CO}$ for each subsystem by first eliminating $P_i(c)$ from the local behavior $L(M_i^{CO})$ (line 11) and then computing the supremal prefix-closed sublanguage of the resulting language (line 12) for each $i\in I$. The candidate coordination supervisor $S_i^{CO}$ is synthesized with respect to the updated $L_i^{CO}$ accordingly (lines 13 and 14). Finally, we update the component module for each subsystem in such a way that $L(M_i^{CO})=P_i^F\left[L(S_i^{CO}\|S_i^F\|G_i^F)\right]$ and $L(M_j^{CO})=L(S_j^{CO}\|G_j)$ if $j\ne i$, respectively (lines 15 to 17). The updated component modules are returned to the compositional verification to determine whether or not $\|_{i\in I} M_i^{CO}\models L$ until no more counterexample is generated. The following theorem ensures the correctness and termination of the SYN-CO algorithm.

\begin{theorem}\label{synco}
Given the component modules $M_i^F$ and $M_j$ $(j\ne i)$, the SYN-CO algorithm terminates and correctly returns the coordination supervisors $S_i^{CO}$ $(i\in I)$.
\end{theorem}
\begin{IEEEproof}
The termination of the SYN-CO algorithm holds due to the fact that during each iteration, each of the updated component module $M_i^{CO}$ $(i\in I)$ possesses a finite number of states regardless of possible faults, and the deployment of the coordination supervisor $S_i^{CO}$ introduces a reduction of the number of states.

Furthermore, it has been shown that the compositional verification with the SYM-N proof rule always terminates and correctly reports whether or not $\|_{i\in I} M_i^{CO}\models L$ \cite{puasuareanu2008learning}; in other words, when no more counterexample is generated, we can conclude that $\|_{i\in I} M_i^{CO}\models L$ will not violate the specification $L$, i.e., the correctness of the SYN-CO algorithm can be achieved.
\end{IEEEproof}

\begin{remark}
It is worth pointing out that although Theorem~\ref{synco} ensures a correct post-fault coordination strategy among the subsystems, the SYN-CO algorithm may still come up with a trivial solution, i.e., $\|_{i\in I} M_i^{CO}=\varnothing$, which always solves Problem~\ref{dcccp}. To prevent this situation from emerging, we abandon the trivial solution returned by the SYN-CO algorithm; instead, we inherit the post-fault supervisor(s) $S_i^F$ for $G_i^F$ and the nominal supervisor $S_j$ for $G_j$ for all $j\ne i$. In this case, the collective behaviors of $G$ are given by $\|_{j\in I-\{i\}} M_j\|M_i^F \not\models L$; nevertheless, from Theorem~\ref{thm:ft}, the coordinated system is still tolerable in the sense that local safety of each subsystem is always assured while additional computation of the coordination supervisor is not required for the nominal subsystems.
\end{remark}

\section{A Multi-robot Coordination Example}\label{sec:experiment}

This section demostrates the effectiveness of the proposed fault tolerant coordination and control framework for distributed DESs through a more comprehensive example.
\begin{figure}[h]
\centering
\includegraphics[width=0.30\textwidth]{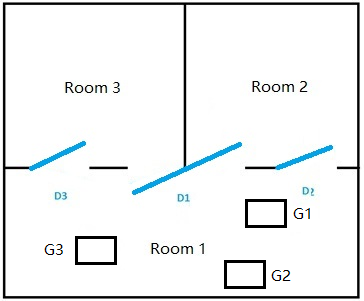}
\caption{The multi-robot coordination scenario.}
\label{scenario}
\vspace{-4.0mm}
\end{figure}

Let us consider a multi-robot system that consists of three mobile robots $G_1$, $G_2$ and $G_3$ with identical localization and communication capabilities. $G_2$ is equipped with fire-extinguishers. All the three robots initially stay in Room 1 in a shared environment as shown in Fig.~\ref{scenario}. Room 2 and Room 1 are connected by the one-way door $D_2$ and the two-way door $D_1$, while Room 3 and Room 1 are connected by $D_1$ and another two-way door $D_3$. $D_1$ is heavy and should be opened by two robots cooperatively. All doors shall close automatically unless there is an external force to keep them open.

To characterize the behaviors of the robots within the DES framework, the events of this example are defined in Table~1.
\begin{table}[t]
  \begin{center}
    \caption{Events of the Multi-robot System}
    \label{list_of_events}
    \begin{tabular}{cc} \toprule
      \multicolumn{1}{l}{Event} &  {Explanation} \\ \midrule
      $h_i$ & Robot $G_i$ receives the service request, $i=1,2,3$.\\
      $G_itoD_1$ & Robot $G_i$ approaches the door $D_1$, $i=1,3$. \\
      $G_ionD_1$ & Robot $G_i$ at the door $D_1$, $i=1,3$. \\
      $G_itok$ & Robot $G_i$ heads for Room $k$, $k=1,2,3$. \\
      $G_iink$ & Robot $G_i$ stays at Room $k$, $k=1,2,3$. \\
      $OP$ & command for moving forward to open $D_1$.  \\
      $CL$ & command for moving backward to close $D_1$. \\
      $D_1open$ & $D_1$ is opened. \\
      $D_1closed$ & $D_1$ is closed. \\
      $r$ & All the robots return to Room 1.\\\bottomrule
    \end{tabular}
    \end{center}
  \vspace{-8.5mm}
\end{table}
The local event set $\Sigma_i$ $(i=1,2,3)$ for robot $G_i$ is defined as follows:
\begin{equation*}
\begin{split}
\Sigma_i=\{& h_i,G_itoD_1,G_ionD_1,OP,CL,G_2in1,G_ito3,\\
		   & G_iin3,D_1closed,D_1open,G_ito1,G_iin1,r\}, i=1,3;\\
\Sigma_i=\{& h_2,G_2to2,G_2in2,D_1open,G_2to1,G_2in1,r\}, i=2.
\end{split}
\end{equation*}
We assume that $G_2in1\in\Sigma_1$, $G_2in1\in\Sigma_3$ and $D_1open\in\Sigma_2$ since they all can be viewed as the information that is transmitted among $G_1$, $G_2$ and $G_3$ via communication. All the events are assumed to be locally observable in the nominal mode. The set of each robot's sensor readings is given by $\Sigma_{i,s}:=\{h_i,G_2in1,G_iin3,G_iin1\}$ for $i=1,3$ and $\Sigma_{2,s}:=\{h_2,G_2in2,D_1open,G_2in1\}$. We also define $\Sigma_{i,a}=\Sigma_{i,c}:=\{G_itoD_1,G_ionD_1,OP,CL,G_ito3,G_ito1,r\}$ as the actuators for $i=1,3$, and let $\Sigma_{2,a}=\Sigma_{2,c}:=\{G_2to2,$ $G_2to1,r\}$ be the set of actuators of $G_2$.

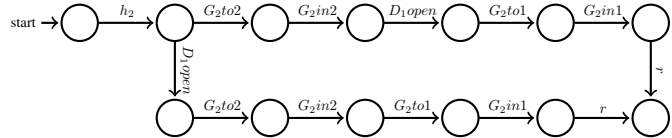
\begin{figure}[h]
\centering
\begin{tikzpicture}[shorten >=1pt,node distance=2.3cm,on grid,auto, bend angle=20, thick,scale=0.55, every node/.style={transform shape}]
	\node[state,initial] (s_0)   {};
	\node[state] (s_1) [right=of s_0] {};
	\node[state] (s_2) [right=of s_1] {};
    \node[state] (s_3) [right=of s_2] {};
	\node[state] (s_4) [right=of s_3] {};
	\node[state] (s_5) [right=of s_4] {};
    \node[state] (s_6) [right=of s_5] {};
	\node[state] (s_7) [below=of s_1] {};
	\node[state] (s_8) [right=of s_7] {};
	\node[state] (s_9) [right=of s_8] {};
	\node[state] (s_10) [right=of s_9] {};
	\node[state] (s_11) [right=of s_10] {};
	\node[state] (s_12) [right=of s_11] {};
	\path[->]
	(s_0) edge node [pos=0.5, sloped, above]{$h_2$} (s_1)
	(s_1) edge node [pos=0.5, sloped, above]{$G_2to2$} (s_2)
	(s_2) edge node [pos=0.5, sloped, above]{$G_2in2$} (s_3)
    (s_3) edge node [pos=0.5, sloped, above]{$D_1open$} (s_4)
	(s_4) edge node [pos=0.5, sloped, above]{$G_2to1$} (s_5)	
	(s_5) edge node [pos=0.5, sloped, above]{$G_2in1$} (s_6)
    (s_1) edge node [pos=0.5, sloped, above]{$D_1open$} (s_7)
	(s_7) edge node [pos=0.5, sloped, above]{$G_2to2$} (s_8)
	(s_8) edge node [pos=0.5, sloped, above]{$G_2in2$} (s_9)
	(s_9) edge node [pos=0.5, sloped, above]{$G_2to1$} (s_10)
	(s_10) edge node [pos=0.5, sloped, above]{$G_2in1$} (s_11)
	(s_11) edge node [pos=0.5, sloped, above]{$r$} (s_12)
    (s_6) edge node [pos=0.5, sloped, above]{$r$} (s_12);
		\end{tikzpicture}
        \caption{The DFA model of the robot $G_2$.}
	    \label{fig:g2plant}
\vspace{-3.5mm}
\end{figure}

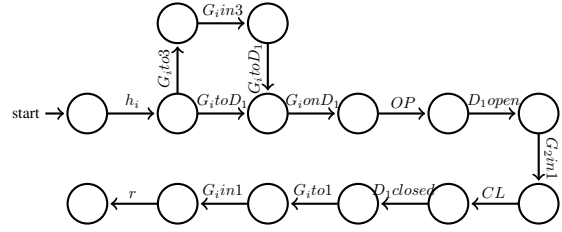
\begin{figure}[h]
\centering
\begin{tikzpicture}[shorten >=1pt,node distance=2.0cm,on grid,auto, bend angle=20, thick,scale=0.60, every node/.style={transform shape}]
	\node[state,initial] (s_0)   {};
		\node[state] (s_1) [right=of s_0] {};
		\node[state] (s_2) [right=of s_1] {};
		\node[state] (s_3) [right=of s_2] {};
		\node[state] (s_4) [right=of s_3] {};
		\node[state] (s_5) [right=of s_4] {};
		\node[state] (s_6) [below=of s_5] {};
        \node[state] (s_7) [left=of s_6] {};
        \node[state] (s_8) [left=of s_7] {};
		\node[state] (s_9) [left=of s_8] {};
		\node[state] (s_10) [left=of s_9] {};
	    \node[state] (s_11) [left=of s_10] {};
        \node[state] (s_12) [above=of s_1] {};
        \node[state] (s_13) [right=of s_12] {};
		\path[->]
		(s_0) edge node [pos=0.5, sloped, above]{$h_i$} (s_1)
		(s_1) edge node [pos=0.5, sloped, above]{$G_itoD_1$} (s_2)
        (s_1) edge node [pos=0.5, sloped, above]{$G_ito3$} (s_12)
		(s_2) edge node [pos=0.5, sloped, above]{$G_ionD_1$} (s_3)
		(s_3) edge node [pos=0.5, sloped, above]{$OP$} (s_4)	
		(s_4) edge node [pos=0.5, sloped, above]{$D_1open$} (s_5)
		(s_5) edge node [pos=0.5, sloped, above]{$G_2in1$} (s_6)
		(s_6) edge node [pos=0.5, sloped, above]{$CL$} (s_7)
		(s_7) edge node [pos=0.5, sloped, above]{$D_1closed$} (s_8)
		(s_8) edge node [pos=0.5, sloped, above]{$G_ito1$} (s_9)
		(s_9) edge node [pos=0.5, sloped, above]{$G_iin1$} (s_10)
        (s_10) edge node [pos=0.5, sloped, above]{$r$} (s_11)
        (s_12) edge node [pos=0.5, sloped, above]{$G_iin3$} (s_13)
        (s_13) edge node [pos=0.5, sloped, above]{$G_itoD_1$} (s_2);
		\end{tikzpicture}
        \caption{The DFA model of the robot $G_i$ $(i=1,3)$.}
	    \label{fig:g13plant}
\vspace{-4.5mm}
\end{figure}

We denote by $G_i$ $(i\in\{1,2,3\})$ the DFA model of $G_i$'s behaviors among the rooms of interest in the environment. Starting from Room 1, $G_2$ can enter Room 2 through $D_2$ and can also move to Room 3 through $D_3$. When $D_1$ is open by the other two robots, $G_2$ can move to both Rooms 2 and 3 through $D_1$. In this example, we only consider the possible behaviors of $G_2$ between Rooms 1 to Room 2 and the corresponding model $G_2$ is depicted in Fig.~\ref{fig:g2plant}.
Similarly, we are interested in the motion behaviors of $G_i$ $(i=1,3)$ between Rooms 1 and 3 of the robot $G_i$ is shown in Fig.~\ref{fig:g13plant}. The model of the distributed multi-robot system is then obtained as $G=G_1\|G_2\|G_3$. In this example, we assume all the motions of each robot within the given environment as safety behaviors; that is, $L_i^{safe}=L(G_i)$ $(i\in\{1,2,3\})$.

\begin{figure}[H]
	\centering
	\begin{tikzpicture}[shorten >=1pt,node distance=2.1cm,on grid,auto, bend angle=20, thick,scale=0.55, every node/.style={transform shape}]
	\node[state,initial] (s_0)   {};
\node[state] (s_1) [right=of s_0] {};
\node[state] (s_2) [right=of s_1] {};
\node[state] (s_3) [right=of s_2] {};
\node[state] (s_4) [right=of s_3] {};
\node[state] (s_5) [right=of s_4] {};
\node[state] (s_6) [right=of s_5] {};
\node[state] (s_7) [below=of s_6] {};
\node[state] (s_8) [left=of s_7] {};
\node[state] (s_9) [left=of s_8] {};
\node[state] (s_10)[left=of s_9] {};
\node[state] (s_11)[left=of s_10] {};
\node[state] (s_12)[left=of s_11] {};
\node[state] (s_13)[left=of s_12] {};
\path[->]
(s_0) edge node [pos=0.5, sloped, above]{$h_3$} (s_1)
(s_1) edge node [pos=0.5, sloped, above]{$G_3to3$} (s_2)
(s_2) edge node [pos=0.5, sloped, above]{$G_3in3$} (s_3)
(s_3) edge node [pos=0.5, sloped, above]{$G_3toD_1$} (s_4)	
(s_4) edge node [pos=0.5, sloped, above]{$G_3onD_1$} (s_5)
(s_5) edge node [pos=0.5, sloped, above]{$OP$} (s_6)
(s_6) edge node [pos=0.5, sloped, above]{$D_1open$} (s_7)
(s_7) edge node [pos=0.5, sloped, above]{$G_2in1$} (s_8)
(s_8) edge node [pos=0.5, sloped, above]{$CL$} (s_9)
(s_9) edge node [pos=0.5, sloped, above]{$D_1closed$} (s_10)
(s_10) edge node [pos=0.5, sloped, above]{$G_3to1$} (s_11)
(s_11) edge node [pos=0.5, sloped, above]{$G_3in1$} (s_12)
(s_12) edge node [pos=0.5, sloped, above]{$r$} (s_13);
	\end{tikzpicture}
	\caption{The nominal supervisor $S_3$.}
	\label{fig:S3}
\vspace{-4.5mm}
\end{figure}
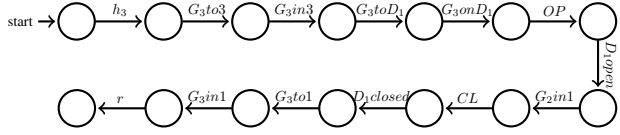

We assume that a fire alarm is triggered in Room 2. The goal is that $G_2$ must respond promptly to the alarm by entering Room 2 through $D_2$ and then return to Room 1. Since $D_2$ is a one-way door and cannot open in Room 2, $G_1$ and $G_3$ need to open $D_1$ jointly so that $G_2$ can successfully return. After that, $G_1$ and $G_3$ should close $D_1$ and return to Room 1 as well. Such a global specification $L$ consists of the following local specifications $L_i^{spe}$ $(i\in\{1,2,3\})$ for $G_i$.
\begin{equation*}
\begin{split}
L_1^{spe}=&\overline{h_1G_1toD_1G_1onD_1OPD_1openG_2in1CL D_1closed}\\
&\overline{G_1to1G_1in1r};\\
L_2^{spe}=&\overline{h_2G_2to2G_2in2D_1openG_2to1G_2in1r}; \\
L_3^{spe}=&\overline{h_3G_3to3G_3in3G_3toD_1 G_3onD_1OPD_1openG_2in1} \\
& \overline{CLD_1closedG_3to1G_3in1r}
\end{split}
\end{equation*}
The global specification $L$ is then given by $L=L^{spe}_1\| L^{spe}_2\|$ $L_3^{spe}$. To satisfy $L$ jointly, on the one hand, robot $G_1$ stays in Room 1 while $G_3$ goes to Room 3 in order to open $D_1$. On the other hand, $G_2$ should enter Room 2 to extinguish the fire and then return to Room 1 as long as $D_1$ is open. The nominal supervisors $S_1$, $S_2$ and $S_3$ are illustrated in Fig.~\ref{fig:S1}, Fig.~\ref{fig:S2} and Fig.~\ref{fig:S3}, respectively.
\begin{figure}[h]
	\centering
\begin{tikzpicture}[shorten >=1pt,node distance=2.2cm,on grid,auto, bend angle=20, thick,scale=0.60, every node/.style={transform shape}]
		\node[state,initial] (s_0)   {};
		\node[state] (s_1) [right=of s_0] {};
		\node[state] (s_2) [right=of s_1] {};
		\node[state] (s_3) [right=of s_2] {};
		\node[state] (s_4) [right=of s_3] {};
		\node[state] (s_5) [right=of s_4] {};
		\node[state] (s_6) [below=of s_5] {};
		\node[state] (s_7) [left=of s_6] {};
		\node[state] (s_8) [left=of s_7] {};
        \node[state] (s_9) [left=of s_8] {};
        \node[state] (s_10) [left=of s_9] {};
        \node[state] (s_11) [left=of s_10] {};
		\path[->]
		(s_0) edge node [pos=0.5, sloped, above]{$h_1$} (s_1)
		(s_1) edge node [pos=0.5, sloped, above]{$G_1toD_1$} (s_2)
		(s_2) edge node [pos=0.5, sloped, above]{$G_1onD_1$} (s_3)
		(s_3) edge node [pos=0.5, sloped, above]{$OP$} (s_4)	
		(s_4) edge node [pos=0.5, sloped, above]{$D_1open$} (s_5)
		(s_5) edge node [pos=0.5, sloped, above]{$G_2in1$} (s_6)
		(s_6) edge node [pos=0.5, sloped, above]{$CL$} (s_7)
		(s_7) edge node [pos=0.5, sloped, above]{$D_1closed$} (s_8)
		(s_8) edge node [pos=0.5, sloped, above]{$G_ito1$} (s_9)
        (s_9) edge node [pos=0.5, sloped, above]{$G_1in1$} (s_10)
        (s_10) edge node [pos=0.5, sloped, above]{$r$} (s_11);    	
		\end{tikzpicture}
        \caption{The nominal supervisor $S_1$.}
	    \label{fig:S1}
\vspace{-4.5mm}
\end{figure}
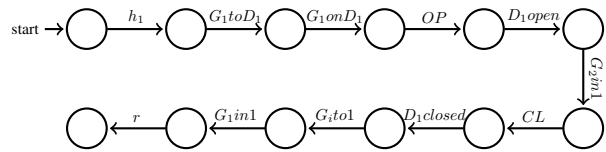

\begin{figure}[h]
	\centering
	\begin{tikzpicture}[shorten >=1pt,node distance=2.2cm,on grid,auto, bend angle=20, thick,scale=0.60, every node/.style={transform shape}]
	\node[state,initial] (s_0)   {};
	\node[state] (s_1) [right=of s_0] {};
	\node[state] (s_2) [right=of s_1] {};
	\node[state] (s_3) [right=of s_2] {};
	\node[state] (s_4) [below=of s_3] {};
	\node[state] (s_5) [left=of s_4] {};
	\node[state] (s_6) [left=of s_5] {};
	\node[state] (s_7) [left=of s_6] {};
	
	\path[->]
	(s_0) edge node [pos=0.5, sloped, above]{$h_2$} (s_1)
	(s_1) edge node [pos=0.5, sloped, above]{$G_2to2$} (s_2)
	(s_2) edge node [pos=0.5, sloped, above]{$G_2in2$} (s_3)
    (s_3) edge node [pos=0.5, sloped, above]{$D_1open$} (s_4)
	(s_4) edge node [pos=0.5, sloped, above]{$G_2to1$} (s_5)	
	(s_5) edge node [pos=0.5, sloped, above]{$G_2in1$} (s_6)
	(s_6) edge node [pos=0.5, sloped, above]{$r$} (s_7);
	\end{tikzpicture}
	\caption{The nominal supervisor $S_2$.}
	\label{fig:S2}
    \vspace{-3.5mm}
    \end{figure}
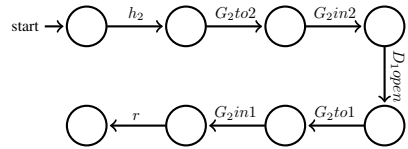

It is worth pointing out that Fig.~\ref{fig:S1}, Fig.~\ref{fig:S2} and Fig.~\ref{fig:S3} also demonstrate the controlled subsystems $G_1^0$, $G_2^0$ and $G_3^0$ in the nominal mode, respectively; therefore, $L(S_i)=L_i$.

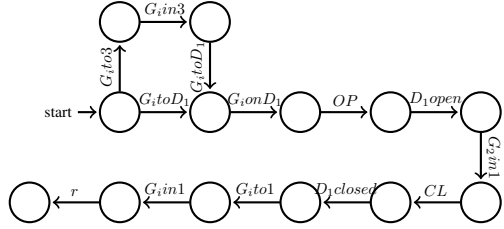
\begin{figure}[h]
\centering
\begin{tikzpicture}[shorten >=1pt,node distance=2.0cm,on grid,auto, bend angle=20, thick,scale=0.60, every node/.style={transform shape}]
	\node[state,initial] (s_1)   {};
		\node[state] (s_2) [right=of s_1] {};
		\node[state] (s_3) [right=of s_2] {};
		\node[state] (s_4) [right=of s_3] {};
		\node[state] (s_5) [right=of s_4] {};
		\node[state] (s_6) [below=of s_5] {};
        \node[state] (s_7) [left=of s_6] {};
        \node[state] (s_8) [left=of s_7] {};
		\node[state] (s_9) [left=of s_8] {};
		\node[state] (s_10) [left=of s_9] {};
	    \node[state] (s_11) [left=of s_10] {};
        \node[state] (s_12) [above=of s_1] {};
        \node[state] (s_13) [right=of s_12] {};
		\path[->]
		(s_1) edge node [pos=0.5, sloped, above]{$G_itoD_1$} (s_2)
        (s_1) edge node [pos=0.5, sloped, above]{$G_ito3$} (s_12)
		(s_2) edge node [pos=0.5, sloped, above]{$G_ionD_1$} (s_3)
		(s_3) edge node [pos=0.5, sloped, above]{$OP$} (s_4)	
		(s_4) edge node [pos=0.5, sloped, above]{$D_1open$} (s_5)
		(s_5) edge node [pos=0.5, sloped, above]{$G_2in1$} (s_6)
		(s_6) edge node [pos=0.5, sloped, above]{$CL$} (s_7)
		(s_7) edge node [pos=0.5, sloped, above]{$D_1closed$} (s_8)
		(s_8) edge node [pos=0.5, sloped, above]{$G_ito1$} (s_9)
		(s_9) edge node [pos=0.5, sloped, above]{$G_iin1$} (s_10)
        (s_10) edge node [pos=0.5, sloped, above]{$r$} (s_11)
        (s_12) edge node [pos=0.5, sloped, above]{$G_iin3$} (s_13)
        (s_13) edge node [pos=0.5, sloped, above]{$G_itoD_1$} (s_2);
		\end{tikzpicture}
        \caption{The DFA model of the robot $G_i$ $(i=1,3)$.}
	    \label{fig:g13af}
\vspace{-3.5mm}
\end{figure}

We now study the local fault tolerant control and post-fault coordination of the multi-robot system in the presence of potential actuator/sensor faults. Herein, we first consider the case in which after receiving the task request $h_3$, $G_3$ is unable to go to Room 3 to open $D_1$ cooperatively with $G_1$. Such a circumstance may correspond to the shortage of battery of $G_3$ and can be captured by the controllability loss of the actuator event $G_3toD_1$. Since the corresponding actuator fault event $h_{3,G_3toD_1}$ occurs after the execution of $h_3$, the post-fault model of $G_3$ is given by $G_3^{F,a}=G_3^{suf}(h_3)$ and is depicted in Fig.~\ref{fig:g13af}.

\begin{figure}[h]
	\centering
\begin{tikzpicture}[shorten >=1pt,node distance=2.2cm,on grid,auto, bend angle=20, thick,scale=0.60, every node/.style={transform shape}]
		\node[state,initial] (s_1)   {};
		\node[state] (s_2) [right=of s_1] {};
		\node[state] (s_3) [right=of s_2] {};
		\node[state] (s_4) [right=of s_3] {};
		\node[state] (s_5) [right=of s_4] {};
		\node[state] (s_6) [below=of s_5] {};
		\node[state] (s_7) [left=of s_6] {};
		\node[state] (s_8) [left=of s_7] {};
        \node[state] (s_9) [left=of s_8] {};
        \node[state] (s_10) [left=of s_9] {};
        \node[state] (s_11) [left=of s_10] {};
		\path[->]
		(s_1) edge node [pos=0.5, sloped, above]{$G_3toD_1$} (s_2)
		(s_2) edge node [pos=0.5, sloped, above]{$G_3onD_1$} (s_3)
		(s_3) edge node [pos=0.5, sloped, above]{$OP$} (s_4)	
		(s_4) edge node [pos=0.5, sloped, above]{$D_1open$} (s_5)
		(s_5) edge node [pos=0.5, sloped, above]{$G_2in1$} (s_6)
		(s_6) edge node [pos=0.5, sloped, above]{$CL$} (s_7)
		(s_7) edge node [pos=0.5, sloped, above]{$D_1closed$} (s_8)
		(s_8) edge node [pos=0.5, sloped, above]{$G_3to1$} (s_9)
        (s_9) edge node [pos=0.5, sloped, above]{$G_3in1$} (s_10)
        (s_10) edge node [pos=0.5, sloped, above]{$r$} (s_11);    	
		\end{tikzpicture}
        \caption{The post-fault supervisor $S_3^{F,a}$.}
	    \label{fig:S3FA}
\vspace{-4.5mm}
\end{figure}
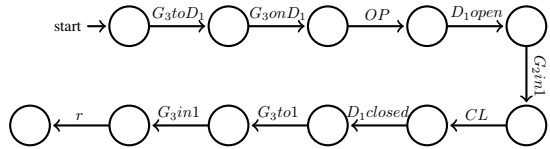

The actuator fault tolerant architecture proposed in Section~\ref{sec:actuator} is utilized at this point to synthesize the post-fault supervisor $S_3^{F,a}$. In this example, the post-fault specification is given by $L_3^{post}=L_3^{safe}/h_3=L(G_3)/h_3$, and it can be verified that the necessary and sufficient conditions presented in Theorem~\ref{atolerance} can be satisfied. Therefore, there exists a satisfactory post-fault supervisor $S_3^{F,a}$, whose DFA representation is shown in Fig.~\ref{fig:S3FA}, that enforces the local safety requirement $L_3^{safe}=L(G_3)$ in the faulty mode.

\begin{figure}[H]
	\centering
	\begin{tikzpicture}[shorten >=1pt,node distance=2.1cm,on grid,auto, bend angle=20, thick,scale=0.55, every node/.style={transform shape}]
	\node[state,initial] (s_1)   {};
\node[state] (s_1) [right=of s_0] {};
\node[state] (s_2) [right=of s_1] {};
\node[state] (s_3) [right=of s_2] {};
\node[state] (s_4) [right=of s_3] {};
\node[state] (s_5) [right=of s_4] {};
\node[state] (s_6) [right=of s_5] {};
\node[state] (s_7) [below=of s_6] {};
\node[state] (s_8) [left=of s_7] {};
\node[state] (s_9) [left=of s_8] {};
\node[state] (s_10)[left=of s_9] {};
\node[state] (s_11)[left=of s_10] {};
\node[state] (s_12)[left=of s_11] {};
\node[state] (s_13)[left=of s_12] {};
\path[->]
(s_0) edge node [pos=0.5, sloped, above]{$h_1$} (s_1)
(s_1) edge node [pos=0.5, sloped, above]{$G_1to3$} (s_2)
(s_2) edge node [pos=0.5, sloped, above]{$G_1in3$} (s_3)
(s_3) edge node [pos=0.5, sloped, above]{$G_1toD_1$} (s_4)	
(s_4) edge node [pos=0.5, sloped, above]{$G_1onD_1$} (s_5)
(s_5) edge node [pos=0.5, sloped, above]{$OP$} (s_6)
(s_6) edge node [pos=0.5, sloped, above]{$D_1open$} (s_7)
(s_7) edge node [pos=0.5, sloped, above]{$G_2in1$} (s_8)
(s_8) edge node [pos=0.5, sloped, above]{$CL$} (s_9)
(s_9) edge node [pos=0.5, sloped, above]{$D_1closed$} (s_10)
(s_10) edge node [pos=0.5, sloped, above]{$G_1to1$} (s_11)
(s_11) edge node [pos=0.5, sloped, above]{$G_1in1$} (s_12)
(s_12) edge node [pos=0.5, sloped, above]{$r$} (s_13);
	\end{tikzpicture}
	\caption{The coordination supervisor $S_1^{CO}$.}
	\label{fig:S1CO}
\vspace{-4.5mm}
\end{figure}
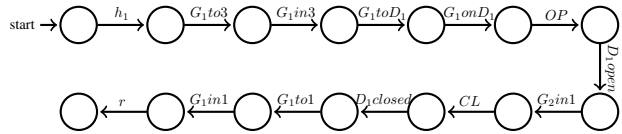

Under the supervision of $S_3^{F,a}$, the robot $G_3$ should stays in Room 1 and open $D_1$. After the post-fault supervisor reconfiguration, the assume-guarantee coordination framework shown in Fig.~\ref{agfmulti} is applied to coordinate $G_3$ with the other two robots. In this case, we can verify that the Proposition~\ref{limif} cannot hold any more and we need to design the coordination supervisors for each subsystem. In particular, it is easy to find out that $G_2$ need not reconfigure the local control policies and $S_2^{CO}=S_2$, whereas the coordination supervisor $S_1^{CO}$ can be synthesized for the robot $G_1$, as shown in Fig.~\ref{fig:S1CO}

In other words, the robot $G_1$ should enter Room 3 to open $D_1$ cooperatively with $G_3$ so that the global task can be achieved.

\section{Conclusion and Future Work}\label{sec:conclusion}
In this paper, we present a fault tolerant coordination and control framework for distributed DESs that are composed of multiple subsystems. The proposed coordination and control framework ensures the accomplishment of the global specification in the presence of sensor and actuator faults. By introducing automata-theoretic methods to characterize the behaviors of each subsystem that is affected by various faults, appropriate post-fault supervisors are synthesized such that local safety can be ensured. In addition, an assume-guarantee coordination scheme is exploited to accomplish the global specification after the post-fault supervisor reconfiguration. The effectiveness of our proposed approach is demonstrated by an illustrative example.

Some problems are of interest for future investigations. For example, extensions of the proposed fault tolerant coordination and control architecture to combinations of other types of faults are worth of study. In addition to fault tolerant control problems, coordination and control reconfiguration strategies for distributed DESs subject to intentional attacks are also expected to be explored.


\bibliographystyle{IEEEtran}        
\bibliography{fault_tolerance}

\begin{thebibliography}{10}
\providecommand{\url}[1]{#1}
\csname url@samestyle\endcsname
\providecommand{\newblock}{\relax}
\providecommand{\bibinfo}[2]{#2}
\providecommand{\BIBentrySTDinterwordspacing}{\spaceskip=0pt\relax}
\providecommand{\BIBentryALTinterwordstretchfactor}{4}
\providecommand{\BIBentryALTinterwordspacing}{\spaceskip=\fontdimen2\font plus
\BIBentryALTinterwordstretchfactor\fontdimen3\font minus
  \fontdimen4\font\relax}
\providecommand{\BIBforeignlanguage}[2]{{%
\expandafter\ifx\csname l@#1\endcsname\relax
\typeout{** WARNING: IEEEtran.bst: No hyphenation pattern has been}%
\typeout{** loaded for the language `#1'. Using the pattern for}%
\typeout{** the default language instead.}%
\else
\language=\csname l@#1\endcsname
\fi
#2}}
\providecommand{\BIBdecl}{\relax}
\BIBdecl

\bibitem{cassandras2008introduction}
C.~G. Cassandras and S.~Lafortune, \emph{Introduction to Discrete Event
  Systems}, 2nd~ed.\hskip 1em plus 0.5em minus 0.4em\relax New York: Springer,
  2008.

\bibitem{blanke2006diagnosis}
M.~Blanke, M.~Kinnaert, J.~Lunze, and M.~Staroswiecki, \emph{Diagnosis and
  Fault-Tolerant Control}, 2nd~ed.\hskip 1em plus 0.5em minus 0.4em\relax
  Berlin: Springer, 2006.

\bibitem{sampath1995diagnosability}
M.~Sampath, R.~Sengupta, S.~Lafortune, K.~Sinnamohideen, and D.~Teneketzis,
  ``Diagnosability of discrete-event systems,'' \emph{IEEE Trans. Autom.
  Control}, vol.~40, no.~9, pp. 1555--1575, 1995.

\bibitem{contant2006diagnosability}
O.~Contant, S.~Lafortune, and D.~Teneketzis, ``Diagnosability of discrete event
  systems with modular structure,'' \emph{Discrete Event Dynam. Syst.: Theory
  Applicat.}, vol.~16, no.~1, pp. 9--37, 2006.

\bibitem{qiu2006decentralized}
W.~Qiu and R.~Kumar, ``Decentralized failure diagnosis of discrete event
  systems,'' \emph{IEEE Trans. Syst., Man, Cybern. A, Syst.,Humans}, vol.~36,
  no.~2, pp. 384--395, 2006.

\bibitem{su2005global}
R.~Su and W.~M. Wonham, ``Global and local consistencies in distributed fault
  diagnosis for discrete-event systems,'' \emph{IEEE Trans. Autom. Control},
  vol.~50, no.~12, pp. 1923--1935, 2005.

\bibitem{schmidt2013verification}
K.~W. Schmidt, ``Verification of modular diagnosability with local
  specifications for discrete-event systems,'' \emph{IEEE Trans. Syst., Man,
  Cybern., Syst.}, vol.~43, no.~5, pp. 1130--1140, 2013.

\bibitem{zaytoon2013overview}
J.~Zaytoon and S.~Lafortune, ``Overview of fault diagnosis methods for discrete
  event systems,'' \emph{Annu. Rev. Control}, vol.~37, no.~2, pp. 308--320,
  2013.

\bibitem{lafortune1991tolerable}
S.~Lafortune and F.~Lin, ``On tolerable and desirable behaviors in supervisory
  control of discrete event systems,'' \emph{Discrete Event Dynam. Syst.:
  Theory Applicat.}, vol.~1, no.~1, pp. 61--92, 1991.

\bibitem{iordache2004resilience}
M.~V. Iordache and P.~J. Antsaklis, ``Resilience to failures and
  reconfigurations in the supervision based on place invariants,'' in
  \emph{Proc. 2004 Amer. Control Conf. (ACC)}, 2004, pp. 4477--4482.

\bibitem{lin1993robust}
F.~Lin, ``Robust and adaptive supervisory control of discrete event systems,''
  \emph{IEEE Trans. Autom. Control}, vol.~38, no.~12, pp. 1848--1852, 1993.

\bibitem{rohloff2005sensor}
K.~R. Rohloff, ``Sensor failure tolerant supervisory control,'' in \emph{Proc.
  44th IEEE Conf. Decision Control Eur. Control Conf. (CDC-ECC)}.\hskip 1em
  plus 0.5em minus 0.4em\relax IEEE, 2005, pp. 3493--3498.

\bibitem{sanchez2006safe}
A.~M. S{\'a}nchez and F.~J. Montoya, ``Safe supervisory control under
  observability failure,'' \emph{Discrete Event Dynam. Syst.: Theory
  Applicat.}, vol.~16, no.~4, pp. 493--525, 2006.

\bibitem{sulek2013computation}
A.~N. S{\"u}lek and K.~W. Schmidt, ``Computation of fault-tolerant supervisors
  for discrete event systems,'' \emph{IFAC Proceedings Volumes}, vol.~46,
  no.~22, pp. 115--120, 2013.

\bibitem{wen2008framework}
Q.~Wen, R.~Kumar, J.~Huang, and H.~Liu, ``A framework for fault-tolerant
  control of discrete event systems,'' \emph{IEEE Trans. Autom. Control},
  vol.~53, no.~8, pp. 1839--1849, 2008.

\bibitem{karimadini2011fault}
M.~Karimadini and H.~Lin, ``Fault-tolerant cooperative tasking for multi-agent
  systems,'' \emph{Int. J. Control}, vol.~84, no.~12, pp. 2092--2107, 2011.

\bibitem{takai2000reliable}
S.~Takai and T.~Ushio, ``Reliable decentralized supervisory control of discrete
  event systems,'' \emph{IEEE Trans. Syst., Man, Cybern., B, Cybern.}, vol.~30,
  no.~5, pp. 661--667, 2000.

\bibitem{liu2010reliable}
F.~Liu and H.~Lin, ``Reliable supervisory control for general architecture of
  decentralized discrete event systems,'' \emph{Automatica}, vol.~46, no.~9,
  pp. 1510--1516, 2010.

\bibitem{kumar2012framework}
R.~Kumar and S.~Takai, ``A framework for control-reconfiguration following
  fault-detection in discrete event systems,'' \emph{IFAC Proceedings Volumes},
  vol.~45, no.~20, pp. 848--853, 2012.

\bibitem{darabi2003control}
H.~Darabi, M.~A. Jafari, and A.~L. Buczak, ``A control switching theory for
  supervisory control of discrete event systems,'' \emph{IEEE Trans. Robot.
  Autom.}, vol.~19, no.~1, pp. 131--137, 2003.

\bibitem{paoli2011active}
A.~Paoli, M.~Sartini, and S.~Lafortune, ``Active fault tolerant control of
  discrete event systems using online diagnostics,'' \emph{Automatica},
  vol.~47, no.~4, pp. 639--649, 2011.

\bibitem{shu2014fault}
S.~Shu and F.~Lin, ``Fault-tolerant control for safety of discrete-event
  systems,'' \emph{IEEE Trans. Autom. Sci. Eng.}, vol.~11, no.~1, pp. 78--89,
  2014.

\bibitem{cobleigh2003learning}
J.~M. Cobleigh, D.~Giannakopoulou, and C.~S. P{\u{a}}s{\u{a}}reanu, ``Learning
  assumptions for compositional verification,'' in \emph{Proc. TACAS
  2003}.\hskip 1em plus 0.5em minus 0.4em\relax Springer, 2003, pp. 331--346.

\bibitem{dai2015learning}
J.~Dai and H.~Lin, ``Learning-based design of fault-tolerant cooperative
  multi-agent systems,'' in \emph{Proc. 2015 Amer. Control Conf. (ACC)}.\hskip
  1em plus 0.5em minus 0.4em\relax IEEE, 2015, pp. 1929--1934.

\bibitem{ramadge1987supervisory}
P.~J. Ramadge and W.~M. Wonham, ``Supervisory control of a class of discrete
  event processes,'' \emph{SIAM J. Control Optim.}, vol.~25, no.~1, pp.
  206--230, 1987.

\bibitem{ramadge1989control}
------, ``The control of discrete event systems,'' \emph{Proc. of the IEEE},
  vol.~77, no.~1, pp. 81--98, 1989.

\bibitem{puasuareanu2008learning}
C.~S. P{\u{a}}s{\u{a}}reanu, D.~Giannakopoulou, M.~G. Bobaru, J.~M. Cobleigh,
  and H.~Barringer, ``Learning to divide and conquer: Applying the \protect{L*}
  algorithm to automate assume-guarantee reasoning,'' \emph{Formal Methods
  Syst. Des.}, vol.~32, no.~3, pp. 175--205, 2008.

\bibitem{kumar1995modeling}
R.~Kumar and V.~K. Garg, \emph{Modeling and Control of Logical Discrete Event
  Systems}.\hskip 1em plus 0.5em minus 0.4em\relax Boston: Kluwer, 1995.

\bibitem{yin2016synthesis}
X.~Yin and S.~Lafortune, ``Synthesis of maximally permissive supervisors for
  partially-observed discrete-event systems,'' \emph{IEEE Trans. Autom.
  Control}, vol.~61, no.~5, pp. 1239--1254, 2016.

\bibitem{carvalho2018detection}
L.~K. Carvalho, Y.-C. Wu, R.~Kwong, and S.~Lafortune, ``Detection and
  mitigation of classes of attacks in supervisory control systems,''
  \emph{Automatica}, vol.~97, pp. 121--133, 2018.

\bibitem{carvalho2013robust}
L.~K. Carvalho, M.~V. Moreira, J.~C. Basilio, and S.~Lafortune, ``Robust
  diagnosis of discrete-event systems against permanent loss of observations,''
  \emph{Automatica}, vol.~49, no.~1, pp. 223--231, 2013.

\bibitem{willner1991supervisory}
Y.~Willner and M.~Heymann, ``Supervisory control of concurrent discrete-event
  systems,'' \emph{Int. J. Control}, vol.~54, no.~5, pp. 1143--1169, 1991.

\bibitem{jiang2000decentralized}
S.~Jiang and R.~Kumar, ``Decentralized control of discrete event systems with
  specializations to local control and concurrent systems,'' \emph{IEEE Trans.
  Syst. Man, Cybern. B, Cybern.}, vol.~30, no.~5, pp. 653--660, 2000.

\bibitem{heymann1994line}
M.~Heymann and F.~Lin, ``On-line control of partially observed discrete event
  systems,'' \emph{Discrete Event Dynam. Syst.: Theory Applicat.}, vol.~4,
  no.~3, pp. 221--236, 1994.

\bibitem{hadj1996centralized}
N.~B. Hadj-Alouane, S.~Lafortune, and F.~Lin, ``Centralized and distributed
  algorithms for on-line synthesis of maximal control policies under partial
  observation,'' \emph{Discrete Event Dynam. Syst.: Theory Applicat.}, vol.~6,
  no.~4, pp. 379--427, 1996.

\bibitem{paoli2005safe}
A.~Paoli and S.~Lafortune, ``Safe diagnosability for fault-tolerant supervision
  of discrete-event systems,'' \emph{Automatica}, vol.~41, no.~8, pp.
  1335--1347, 2005.

\end{thebibliography}

\end{document}